%% file: main.tex
\documentclass[a4paper, amsfonts, amssymb, amsmath, reprint, showkeys,showpacs, nofootinbib]{revtex4-2}
\usepackage[english]{babel}
\usepackage[utf8]{inputenc}
\input{preamble}
\usepackage[pdftex, pdftitle={Article}, pdfauthor={Author}]{hyperref} 
\usepackage{graphicx}
\usepackage{float}
\usepackage{hyperref}
\hypersetup{
    colorlinks=true,
    citecolor = blue,
    linkcolor=blue,
    filecolor=magenta,      
    urlcolor=blue}
    
\begin{document}
\title{Proca Stars with Dark Photons from Spontaneous Symmetry Breaking of the Scalar Field Dark Matter}
\author{Leonardo Sán$.$ Hernández$^{1,}$}\email[]{leonardo.sanchez@cinvestav.mx}
\author{Tonatiuh Matos$^{1,}$}\email[]{tonatiuh.matos@cinvestav.mx}
    \affiliation{${ }^{1}$Departamento de Física, Centro de Investigación y de Estudios Avanzados del IPN, A.P. 14-740, 07000 CDMX., México.}
\date{\today} 

\begin{abstract}
Recently, the Scalar Field Dark Matter (SFDM) model (also known as Fuzzy, Wave, Bose-Einstein, Ultra-light
Dark Matter) has gained a lot of attention because it has provided simpler and more natural explanations for various phenomena observed in galaxies, as a natural explanation for the center of galaxies, the number of satellite galaxies around their host and, more recently, a natural explanation for anomalous trajectories of satellite galaxies called Vast Polar Orbits (VPO) observed in various galaxies. In the present work we study the assumption that the SFDM is a type of charged dark boson whose gauge charge is associated with the Dark Photon (DP). Inspired by these results, we study the formation of compact bosonic objects, such as Boson Stars (BS) and focus on the possibility that, due to spontaneous $U(1)$ SFDM symmetry breaking, the DP may acquire mass and form compact objects like Proca Stars (PS). If this is true, we can expect measurable effects on the electromagnetic field of the Standard Model (SM) of particles due to their interaction with the DP on the formation of compact objects.

\end{abstract}

\pacs{95.30.Sf, 95.35.+d, 98.80.Jk}

\maketitle

\input{sections/section01.tex}  
\input{sections/section1.tex}

\input{sections/section02.tex}

\input{sections/SSB.tex}

\input{sections/PS.tex}

\input{sections/PSDP.tex}
\input{sections/Results.tex}

\input{sections/Observables.tex}
\input{sections/section03.tex}

\input{sections/acknowledgements.tex}

\nocite{*}
\bibliography{bib2}


\end{document}

%% file: preamble.tex
\usepackage{amsthm}
\usepackage{mathtools}
\usepackage{physics}
\usepackage{xcolor}
\usepackage{graphicx}
\usepackage[left=23mm,right=13mm,top=35mm,columnsep=15pt]{geometry} 
\usepackage{adjustbox}
\usepackage{placeins}
\usepackage[T1]{fontenc}
\usepackage{lipsum}
\usepackage{csquotes}

%% file: sections/section01.tex
\section{Introduction} \label{sec:outline}
Dark Matter (DM) is one of the most fascinating open problems in physics. The Cold Dark Matter (CDM) model has been one of the most widely used to attempt to explain dark matter; however, this model presents a number of difficulties in making predictions on a galactic scale. Because of this, alternative models that are consistent with the observations have been proposed; the Scalar Field Dark Matter is one of these models.
   
The idea of SFDM model was proposed in 1994 by Ji S. U. $\&$ Sin S. J. \cite{ji_late-time_1994}, and Lee $\&$ Koh in 1996 \cite{lee_galactic_1996}, and later independently by Siddharta $\&$ Matos in 1998 \cite{guzman_scalar_2000}. It was during these years that continued research on scalar field dark matter began. Historical and analytical reviews on the evolution of this model can be found in \cite{magana_brief_2012,urena-lopez_brief_2019,suarez_review_2014,rindler-daller_complex_2014,marsh_warmandfuzzy_2016,niemeyer_small-scale_2020}. The SFDM model has been rediscovered over the years and has been given different names: Fuzzy Dark Matter (2000) \cite{hu_cold_2000}, Quintessential Dark Matter (2001) \cite{arbey_quintessential_2001}, Wave Dark Matter (2010) \cite{bray_wave_nodate} and Ultralight Dark Matter (ULDM), which classifies scalar field dark matter based on its condensation structure \cite{ferreira_ultra-light_2021}.

In general, the SFDM model proposes that dark matter consists of spin-zero ultralight massive bosons associated with a scalar field $\Phi$ that only interacts gravitationally with baryonic matter. So, the SFDM must satisfy an equation of the type Klein-Gordon, and the dynamics of the universe can be found by combining the Lagrangian of this model with gravity in the Einstein equations.

One of the most fascinating characteristics of SFDM is its quantum nature, recently studied in \cite{matos_quantum_2022}. Due to this property, the SFDM model is capable of explaining phenomena such as the vast polar orbits of satellite galaxies around their host, the so-called VPO \cite{solis-lopez_scalar_2021}, observed in systems such as the Milky Way, Andromeda, and Cen A.
The VPOs are explained using excited states of the scalar field. At galactic regimes, non-relativistic behavior and weak fields can be assumed for the system. In this approximation, the Einstein-Klein-Gordon system that describes the SFDM is reduced to a Schr\"odinger-Poisson system, so that the system can be in the ground state or excited states (or both) in analogy to an atom. The excited states produce an 8-shaped structure, which may explain the alignment of the satellite galaxies on polar trajectories around the host galaxies. 
This is a remarkable result as no other DM model can adequately explain these phenomena in such a natural way. 

On the other hand, exotic compact objects, also known as exotic stars or bosonic stars, have also gained great relevance in recent years due to their ability to mimic significant astrophysical phenomena such as Black Hole Shadows (BHS) \cite{herdeiro_imitation_2021, olivares_how_2020, sengo_kerr_2023} and Gravitational Wave Events (GW) \cite{bustillo_gw190521_2021}. This becomes even more important with the discovery of the Higgs boson in 2012, since this was a reaffirmation of the existence of bosons in nature (although the Higgs boson is too unstable to be a candidate to form boson stars \cite {visinelli_boson_2021}). In general, the term Bosonic Star (BSS) refers to hypothetical astrophysical compact objects composed fundamentally of ultralight bosons \cite{herdeiro_imitation_2021}. If the star is composed of bosons with spin $s=0$, it is known as Boson Star (described by scalar fields) \cite{kaup_klein-gordon_1968}, and if it is composed of bosons with spin $s=1$, it is known as Proca Star (described by vector fields) \cite{brito_proca_2016}. Both PS and BS have been studied in the context of DM. In \cite{sharma_boson_2008} BS were studied at galactic scales, and it was found that for scalar bosons of mass $m \sim 10^{-23}$ $eV$ it is possible to obtain stars of mass $ M \sim 10^{12}M_{\odot}$ and radius $R\sim 10^{13}$ $km$. These results are compatible with the SFDM model for the formation of galactic nuclei presented in \cite{ alcubierre_galactic_2002}. On the other hand, PS have been used to make simulations that mimic black hole observables. This is possible because, unlike the BS, the PS possess a robust formational dynamics process \cite{herdeiro_black_2022}. In \cite{bustillo_gw190521_2021}, for example, the gravitational wave event GW190521 was simulated using the collision of two rotating PS, and when the results were compared to the observational data, it was found that this model fits the event a little better than the one proposed by the LIGO-Virgo collaboration \cite{herdeiro_black_2022}. While in \cite{herdeiro_imitation_2021} it was shown that PS are capable of producing shadows similar to those of a Schwarzschild black hole if the observation is made from a polar angle similar to the position of the earth with respect to M87. Then, the capability of PS to mimic dark objects such as black holes motivates the possibility of being studied by DM models. 

Inspired by all these results, in this work we study the possibility of obtaining BS and PS by imposing particular constraints on the SFDM model proposed in \cite{matos_fermi_2022}. In particular, we consider the case where the DP associated with the $U(1)$ gauge symmetry of the model can acquire mass due to the spontaneous symmetry breaking of the system and then form compact objects such as Proca stars. By doing this, we find that the formation of Proca stars has effects on the electromagnetic field of the SM due to the mixing terms between the DP and the SM photon. This is very important because it means that the formation of these compact objects causes changes in the electromagnetic field of the SM that, in principle, we could measure.

In Section \ref{sec:model} we present the general SFDM model that we use and the derivation of some particular cases such as BS and CBS. In Section \ref{sec:SSB}, we present the physical mechanism for the spontaneous symmetry breaking considering that the scalar field is inside a thermal bath, and therefore there is a cut-off term for the temperature that can cause the spontaneous symmetry breaking through which the DP acquires mass. In Section \ref{sec:PS}, we focus on the Proca stars formed by dark photons that acquire mass via the Higgs mechanism induced by the SFDM, and we study the effects of this process on the maximal mass of the Proca stars. In Section \ref{sec:PSDP}, we study the case of Proca stars when the coupling between the DP and SM photon is manifest. Finally, in Sections \ref{sec:Results} and \ref{sec:Observables}, we present the numerical results and the effects of this process on the electromagnetic field of the SM.

%% file: sections/section1.tex
\section{U(1)-SFDM Model} \label{sec:model}
Thus, we start with the Lagrangian proposed in \cite{matos_fermi_2022}
\begin{equation}\label{model}
\begin{aligned}
\mathcal{L} &=-\left(\nabla_{\mu} \Phi+i q B_{\mu} \Phi\right)\left(\nabla^{\mu} \Phi^{*}-i q B^{ \mu} \Phi^{*}\right)-V\left(\Phi\right) \\
&-\frac{1}{4} F_{\mu \nu} F^{\mu \nu}-\frac{1}{4} B_{\mu \nu} B^{\mu \nu}-\frac{\delta^{2}}{2} F_{\mu \nu} B^{\mu \nu},
\end{aligned}
\end{equation}
where $\Phi$ is the complex scalar dark matter field; $B_{\mu}$ is the DP field associated to the $U(1)$ symmetry of the SFDM, with fundamental charge $q$ and Faraday tensor defined as $ B_{\mu \nu}=\nabla_{\mu} B_{\nu}-\nabla_{\nu} B_{\mu}$; $A_{\mu}$ is the $4-$potential of the electromagnetic field of the standard model, with Faraday tensor defined as $ F_{\mu \nu}=\nabla_{\mu} A_{\nu}-\nabla_{\nu} A_{\mu}$; $V\left(\Phi\right)$ is the potential associated to $\Phi$, and $\delta^{2}$ is the kinetic mixing parameter which couples the fields $A_{\mu}$ and $B_{\mu}$. We use the standard general relativity signature $(-, +, +, +)$ in this work and the natural units $\hbar=c=k_{\beta}=\epsilon_{0}=1$. In general, to obtain separable solutions in the study of compact objects, we consider the fields $A_\mu$ and $B_\mu$ to be complex (each can be described in terms of two real fields), so the real part of $A_\mu$ corresponds to the physical electromagnetic field of the SM. Thus we rewrite the Lagrangian (\ref{model}) as follows
\begin{equation}\label{model2}
\begin{aligned}
\mathcal{L} &=-\left(\nabla_{\mu} \Phi+i q B_{\mu} \Phi\right)\left(\nabla^{\mu} \Phi^{*}-i q B^{*\mu} \Phi^{*}\right)-V(\Phi)\\
&-\frac{1}{4} F_{\mu \nu} F^{*\mu \nu}-\frac{1}{4} B_{ \mu \nu} B^{* \mu \nu}-\frac{\delta^{2}}{2} R_{e}\left\{F_{\mu \nu} B^{\mu \nu}\right\}.
\end{aligned}
\end{equation}
It's important to notice that making $A_\mu$ and $B_\mu$ complex introduces one extra real field for each one, so the $U(1)$ symmetry may no longer be present. The reason for proceeding in this way is solely to obtain a time-independent system of differential equations when solving the Einstein-Proca system. However, the $U(1)$ gauge symmetry present in (\ref{model}) emerges only under the particular constraints that $A_\mu$ and $B_\mu$ are real, as is the case for BS, CBS, and SSB. In the case of PS, the complex field approximation must be viewed as two real Proca fields, each with the same mass, that form the real and imaginary parts of the complex vector field representing the spin $s=1$ particle. Therefore, this procedure is valid as long as the correct constraints are applied to preserve the appropriate symmetries in each particular case studied in this work.
We couple this Lagrangian minimally with gravity in an action $S$ of the form
\begin{equation}
    S=\int d^{4} x \sqrt{-g}\left[\frac{1}{16 \pi G} R+\mathcal{L}\right],
\end{equation}
where $R$ is the Ricci tensor, $g$ is the determinant of the metric $g_{\mu\nu}$ and $G$ is the Newton´s constant. By varying this action with respect to the metric, we derive Einstein's field equations $ G_{\alpha \mu}=8 \pi G T_{\alpha \mu}$, where the energy-momentum tensor  reads
\begin{equation}
    \begin{aligned}
T_{\alpha \mu} &=g_{\alpha \mu} \mathcal{L}+\left(\nabla_{\mu} \Phi+i q B_{\mu} \Phi\right)\left(\nabla_{\alpha} \Phi^{*}-i q B_{\alpha}^{*} \Phi^{*}\right) \\
&+\left(\nabla_{\alpha} \Phi+i q B_{\alpha} \Phi\right)\left(\nabla_{\mu} \Phi^{*}-i q B_{\mu}^{*} \Phi^{*}\right)-F_{\nu(\alpha} F_{\mu)}^{* \nu} \\
&-B_{\nu(\alpha} B_{\mu )}^{* \nu}-2\delta^{2} \operatorname{Re}\left[F_{\nu(\alpha} B_{\mu)}^{\ \nu}\right].
\end{aligned}
\end{equation}
The variation of the action $S$ with respect to $\Phi$, $A_{\mu}$, and $B_{\mu}$, gives the following equations of motion, respectively
\begin{equation}\label{KGMD}
    \left(\nabla^{\mu}+i q B^{* \mu}\right)\left(\nabla_{\mu} \Phi+i q B_{\mu} \Phi\right)-\frac{d V\left(\Phi \Phi^{*}\right)}{d|\Phi|^{2}} \Phi=0,
\end{equation}
    \begin{equation}\label{MD}
    \nabla_{\mu} F^{\mu \nu}+\delta^{2} \nabla_{\mu} B^{* \mu \nu}=0,
\end{equation}
 \begin{equation}\label{PMD}
    \nabla_{\mu} B^{\mu \nu}+\delta^{2} \nabla_{\mu} F^{* \mu \nu}=-2 i q \Phi^{*}\left(\nabla^{\nu} \Phi+i q B^{\nu} \Phi\right).
\end{equation}  
As we can see, the equation (\ref{KGMD}) corresponding to the variation of $\Phi$ is a Klein-Gordon-type equation with a modified gauge covariant derivative that makes the complex nature of $B_{\mu}$ manifest. Additionally, the equation (\ref{MD}) corresponding to the variation of $A_{\mu}$ describes the coupled electrodynamics of the SM photon with the DP. Finally, the equation (\ref{PMD}) from the variation of $B_{\mu}$ contains the electrodynamics of the DP coupled with the SM photon, along with the term that makes the coupling between $B_{\mu}$ and $\Phi$ manifest.
We can impose particular constraints on this system to find different cases of compact  objects formed by SFDM bosons. For example, by imposing $A_{\mu}=B_{\mu}=0$ (which corresponds to regimes where electromagnetic fields are negligible) in the Lagrangian (\ref{model2}) we obtain the action $S_{BS}$ associated to scalar boson stars of the form
\begin{equation}
     S_{BS}=\int d^{4} x \sqrt{-g}\left[\frac{R}{16 \pi G}-\nabla_{\mu} \Phi \nabla^{\mu} \Phi^{*}-V\left(\Phi\right)\right].
\end{equation}
In the same way, if we now consider $A_{\mu}=0$ and $B_{\mu}$ real in (\ref{model2}), we find the action $S_{CBS}$ that describes charged boson stars (CBS)
\begin{equation}
    S_{CBS}=\int d^{4} x \sqrt{-g}\left[\frac{R}{16 \pi G} +\mathcal{L}_{CBS}\right],
\end{equation}
where
\begin{equation}
   \begin{aligned}
\mathcal{L}_{C B S} &=-\left(\nabla_{\mu} \Phi+i q B_{\mu} \Phi\right)\left(\nabla^{\mu} \Phi^{*}-i q B^{ \mu} \Phi^{*}\right) \\
&\ -V\left(\Phi\right)-\frac{1}{4} B_{\mu \nu} B^{ \mu \nu}.
\end{aligned}
\end{equation}
In both cases, BS and CBS are formed by SFDM ultralight bosons, and the charge $q$ of the scalar field in CBS is associated with the dark photon. In general, the SFDM model assumes ultralight boson masses between $1-10^{-24}$ $eV$ \cite{ferreira_ultra-light_2021}, whereas models of bosonic stars in the literature assume masses for bosonic particles of approximately $10^{-10}-10^{-20}$ $eV$ to obtain stars of astrophysical interest with maximal masses of approximately $1-10^{10} M_{\odot}$ \cite{herdeiro_black_2022}. Then, we can consider compatible mass ranges to use the boson star literature results to study these compact objects in the context of SFDM presented here. See for example \cite{herdeiro_imitation_2021} for general solutions in BS and \cite{kumar_boson_nodate} for solutions in CBS. 

The particular case that concerns us in this work is to consider a Higgs-type potential for the scalar field of dark matter so that the dark photon field $B_{\mu}$ can acquire mass through an spontaneous symmetry braking (SSB) of the system and then form compact objects like Proca Stars (PS). In the next section, we propose a thermal bath for $\Phi$ as a physical mechanism for the SSB. Thus, there is a range of temperatures and a cutoff temperature for SSB to occur.

%% file: sections/section02.tex
\section{Thermal Bath} \label{sec:develop}
We propose that the complex scalar field of dark matter is in a thermal bath at temperature $T$, which is characterized by the potential of the form \cite{matos_phase_2014, matos_bose-einstein_2017}
\begin{equation}\label{V}
    V=-m_{\Phi}^2 \Phi \Phi^*+\frac{\lambda}{2}\left(\Phi \Phi^*\right)^2+\frac{\lambda}{4} \Phi \Phi^* T^2+\frac{\pi^2}{90} T^4,
\end{equation}
where $m_{\Phi}$ is the mass parameter of the scalar field $\Phi$, and $\lambda$ is the self$-$interaction parameter. In order to study the Higgs mechanism, in this work we just consider the case $\lambda>0$ (repulsive interactions). It is convenient to redefine the zero potential value of this potential and factorize $V$ as follows
\begin{equation}\label{V2}
   V=\frac{\lambda}{2}\left(\phi \phi^*-\frac{m^2}{\lambda}\right)^2,
\end{equation}
where we define the effective mass parameter $m$ of the scalar field as
\begin{equation}\label{m2}
    m^2 \equiv \pm m_\phi^2\left(1-\frac{\lambda}{4 m_\phi^2} T^2\right).
\end{equation}
The plus and minus signs in this definition depend on the sign of the expression between parentheses. To illustrate this, we define the spontaneous symmetry breaking parameter as $\eta \equiv (1-\lambda T^{2} / 4m_\phi^2)$. If $\eta>0$, then $m^2$ is given by the plus sign in (\ref{m2}). In this case the potential (\ref{V2}) describes a Higgs-type potential (Mexican hat shape), and therefore we can expect a SSB for the system. On the other hand, if $\eta\leq 0$, then $m^2$ is given by the minus sign in (\ref{m2}). In this case the Mexican hat shape is lost for (\ref{V2}), so there is no Higgs-type potential and therefore no SSB. We must note that in both cases $m^2>0$, so the effective mass is always real. From this analysis, we define the critical temperature $T_{c}$ for SSB to occur. This temperature is given in terms of $\eta$ as $\eta\left(T_{c}\right)=0$, so that
\begin{equation}
    T_{c}\equiv\frac{2 m_\phi}{\sqrt{\lambda}}.
\end{equation}
The SSB is then possible for temperatures in the range $T \in\left[0, T_{c}\right)$, while for $T \geqslant T_{c}$ there is no SSB. We are only interested in the case where there may be an SSB, so from now on we consider only temperatures $T<T_{c}$. Hence we take the plus sign on (\ref{m2}) and we can rewrite $m^2$ in terms of $T$ and $T_c$ as  
\begin{equation}\label{m2T}
    m^2=m_\phi^2\left(1-\frac{T^2}{T_c^2}\right).
\end{equation}
The associated Higgs potential has a double minimum at $|\Phi|^2 \equiv \Phi \Phi^*=\frac{m^2}{\lambda}$. Then, the expected value of $\Phi$ at the vacuum can be expressed as
\begin{equation}\label{phi0}
    \langle\Phi\rangle_0=\Phi_0=\pm \frac{m}{\sqrt{\lambda}}.
\end{equation}
To obtain a Higgs mechanism via the SSB, we rewrite $\Phi$ as excitations above the vacuum state (\ref{phi0}). Before proceeding, we must note that the vacuum state depends on the effective mass $m$ and hence on the temperature $T$ that we consider for the thermal bath. This is noteworthy because it is this effective mass that we can use to impose constraints in the search for dark matter and comparisons with observables. In the next section, we study the possibility for the DP to obtain mass via a Higgs mechanism induced by the SSB of the ground state (\ref{phi0}). Then, we study the implications of this in the solutions for Proca-type stars.

%% file: sections/SSB.tex
\section{Spontaneous Symmetry Breaking} \label{sec:SSB}
Now, we put the potential (\ref{V2}) into the general Lagrangian (\ref{model2}) of the proposed model and consider temperatures $T<T_{c}$, so that $m$ is given by equation (\ref{m2T}). In this case we can rewrite the complex scalar field $\Phi$ as excitations of the ground state $\Phi_{0}$ (\ref{phi0}) associated to the potential (\ref{V2}). So, using Polar notation, we decompose $\Phi$ in terms of two real scalar fields, $\rho(x)$ and $\theta(x)$, by doing radial perturbations around $\Phi_{0}$ as follows
\begin{equation}
    \Phi=\left[\Phi_0+\rho(x)\right] e^{i \theta(x)}.
\end{equation}
By substituting this expression of $\Phi$ in the general Lagrangian, only the terms of the covariant derivatives and of the potential $V$ are modified, and by choosing the unitary gauge $\theta(x)=0$ to break the symmetry of the ground state, then the Lagrangian $\mathcal{L}$ is now written as
\begin{equation}\label{L2}
\begin{aligned}
\mathcal{L}= & -\frac{1}{4} F_{\mu \nu} F^{* \mu \nu}-\frac{1}{4} B_{\mu \nu} B^{* \mu \nu}-\frac{\delta^2}{2} R_e\left\{F_{\mu \nu} B^{\mu \nu}\right\} \\
& -\nabla^\mu \rho \nabla_\mu \rho-q^2\left(\Phi_0+\rho\right)^2 B_\mu B^{* \mu}-V(\rho),
\end{aligned}
\end{equation}
where
\begin{equation}
    V(\rho)=\frac{\lambda}{2}\left\{4 \Phi_0^2 \rho^2+4 \Phi_0 \rho^3+\rho^4\right\}.
\end{equation}
As we can see, this Lagrangian already contains a mass term $q^{2} \Phi_{0}^{2} B_{\mu} {B}^{* \mu}$ for the DP. So, now we can define the mass $\mu$ of the dark photon field as
\begin{equation}\label{mu}
    \frac{1}{2} \mu^{2} \equiv q^{2} \Phi_{0}^{2}=q^{2} \frac{m^{2}}{\lambda}.
\end{equation}
Therefore, in this model, the mass $\mu$ of the DP is limited by the self-interaction parameter $\lambda$, the effective mass $m$ of the scalar field dark matter, and the gauge charge $q$. The Lagrangian (\ref{L2}) also contains a mass term $2 \lambda \Phi_{0}^{2} \rho^{2}$ for the $\rho$ field. This term is also known as Higgs mode, which, as we saw previously, is associated with radial perturbations of the ground state in the Higgs potential. The Higgs mode has an associated mass defined as $m_{\rho} \equiv \sqrt{2 \lambda} \Phi_{0}=\sqrt{2} m$.
In order to decouple the $\rho$ and $B_{\mu}$ fields and obtain a Proca-like Lagrangian from (\ref{L2}), we could in principle consider a Stueckelberg mechanism, which could be obtained in the case where $\lambda >>q$ (so that the mass $m_{\rho}$ of the Higgs mode is much larger than that of the dark photon $\mu$). This process could be studied in the context of a UV completion theory for the Proca model, so that the Proca model is obtained at low effective energies. However, the Ly$-$alpha observations adjusted for self-interacting SFDM constrain the value of $\lambda$ to be much less than $1$ (of about $10^{-86}$ for ultralight scalar fields with masses of about $m_{\Phi}\sim10^{-22}eV$) in order to satisfy the constraints of nucleosynthesis \cite{li_cosmological_2014}. While, in \cite{matos_fermi_2022}, the production of Fermi bubbles was studied using model (\ref{model}), with values for $q$ around $q\sim10^{-13}$. Therefore, although it is interesting to explore both ranges ($\lambda >q$ and $\lambda <q$), what we can expect is that for SFDM models with ultra-light masses, $\lambda$ is less than $q$. Due to this, the approximation that we use in this work is to consider regions where the scalar dark matter field $\Phi$ is quenched at its expected value in vacuum ($\Phi(x)=\Phi_{0}$), which is equivalent to considering that the radial perturbations of the ground state associated with the Higgs mode are negligible and therefore $\rho(x)=0$. In this case, the Lagrangian (\ref{L2}) can be written simply as
\begin{equation}\label{LDP}
    \begin{aligned}
\mathcal{L}= & -\frac{1}{4} F_{\mu \nu} F^{* \mu \nu}-\frac{1}{4} B_{\mu \nu} B^{* \mu \nu}-\frac{1}{2} \mu^{2} B_{\mu} B^{* \mu} \\
& -\frac{\delta^{2}}{2} R_{e}\left\{F_{\mu \nu} B^{\mu \nu}\right\}.
\end{aligned}
\end{equation}
We should note that although up until now we have considered the case where $A_{\mu}$ and $B_{\mu}$ are complex, the discussion presented is still valid for the case where both are real, and is in this case when we recover the usual SSB process for the DP. So the Lagrangian (\ref{LDP}) simply describes a dark photon with mass $\mu$ and a kinetic mixing parameter $\delta^2$. From this Lagrangian, we can study the formation of Proca stars composed of dark photons, as has been previously done in the general case of spin$-$$1$ bosons. The difference in this case is that the DP acquires mass from its interaction with the SFDM, which imposes constraints on the mass of the dark photon. The approximation studied in this work also has the consequence that in places where there is SFDM, we could find not only massless dark photons, as described in \cite{matos_fermi_2022}, but also dark photons with mass, which can form astrophysically interesting objects.

%% file: sections/PS.tex
\section{Proca Stars} \label{sec:PS}
As a first approximation we consider the case where there is no electromagnetic field of the SM (or it is negligible), so we can set $A_{\mu}=0$. In this case  (\ref{LDP}) reduces to a Proca-type Lagrangian that we can write as
\begin{equation}
    \mathcal{L}_{P}=-\frac{1}{4} B_{\mu \nu} B^{* \mu \nu}-\frac{\mu^{2}}{2} B_{\mu} B^{* \mu}.
\end{equation}
We couple this Lagrangian minimally with gravity and we obtain a Proca-Einstein action $S_{P}$ of the form
\begin{equation}
     S_{P}=\int d^{4} x \sqrt{-g}\left[\frac{R}{16 \pi G}+\mathcal{L}_{P}\right].
\end{equation}
By varying this action with respect to the metric, we find the following Einstein equations
\begin{equation}\label{einstein}
    G_{\mu \nu}=8\pi G\left[g_{\mu \nu} \mathcal{L}_{P}-B_{\alpha(\mu} B_{\nu)}^{* \alpha}+2 \hat{U} B_{(\mu} B_{\nu)}^{*}\right],
\end{equation}
and the variation with respect to $B_{\mu}$ gives Proca's equations of motion
\begin{equation}\label{proca}
    \nabla_{\mu} B^{\mu \nu}=2 \hat{U} B^{\nu},
\end{equation}
where, following the same notation as in \cite{herdeiro_imitation_2021}, we have defined a general potential $U\left(B^{2}\right)=\frac{1}{2} \mu^{2} B_{\mu} B^{* \mu}$ (this notation is useful for the case in which self-interaction quartic potentials are considered for $B_{\mu}$) and $\hat{U} \equiv \frac{d U}{d B^{2}}$, where $ B^{2} \equiv B_{\mu} B^{* \mu}$.
This system has been arduously studied in a general way in \cite{brito_proca_2016} for spherical symmetric and static space-times and in \cite{herdeiro_asymptotically_2019} for the spinning case. In this work, we just consider the spherical symmetric and static case as a first approximation. We proceed in the same manner as in \cite{brito_proca_2016, herdeiro_imitation_2021} and propose a metric of the following form
\begin{equation}\label{metric}
    d S^{2}=-\sigma^{2}(r) N(r) d t^{2}+\frac{d t^{2}}{N(r)}+r^{2} d \Omega_{2},
\end{equation}
and for the dark photon field we propose a Proca-type ansatz of the form
 \begin{equation}\label{Bproca}
      B(r,t)=[f(r) d t+i g(r) d r]e^{-i \omega t},
  \end{equation}
where $N(r)\equiv1-2m(r)/r$; $\sigma(r)$, $m(r)$, $f(r)$ and $g(r)$ are real functions of the radial coordinate $r$, and $\omega$ is a real frequency parameter. We must be careful not to confuse the function $m(r)$ associated with the metric with the effective mass parameter $m$ of the scalar field of dark matter, defined in (\ref{m2T}) (which is a constant). Substituting the metric (\ref{metric}) and the ansatz (\ref{Bproca}) into equations (\ref{einstein}), we obtain that the non-zero Einstein equations are \cite{brito_proca_2016}
\begin{equation}
    m^{\prime}=4 \pi G r^{2}\left[\frac{\left(f^{\prime}-\omega g\right)^{2}}{2 \sigma^{2}}+\frac{\mu^{2}}{2}\left(g^{2} N+\frac{f^{2}}{N \sigma^{2}}\right)\right],
\end{equation}
\begin{equation}
    \frac{\sigma^{\prime}}{\sigma}=4 \pi G r \mu^{2}\left(g^{2}+\frac{f^{2}}{N^{2} \sigma^{2}}\right),
\end{equation}
while the Proca equations (\ref{proca}) are \begin{equation}
    \frac{d}{d r}\left\{\frac{r^{2}\left[f^{\prime}-\omega g\right]}{\sigma}\right\}=\frac{\mu^{2} r^{2} f}{\sigma N},
\end{equation}
\begin{equation}
    \omega g-f^{\prime}=\frac{\mu^{2} \sigma^{2} N g}{\omega},
\end{equation}
where the $^{\prime}$ indicates derivation with respect to the radial coordinate $r$. This system can be solved numerically for the non-asymptotic regions of the system. While for the asymptotic regions ($r \rightarrow \infty$ and $r \rightarrow 0$), it is necessary to do series expansions as in \cite{brito_proca_2016}. The maximal ADM mass $M_{ADM}$ of this type of solution is given in terms of the dark photon mass $\mu$ by the following relation \cite{freitas_ultralight_2021}
\begin{equation}
    M_{A D M}^{\max }=\alpha_{B S} \frac{M_{P l}^2}{\mu}=\alpha_{B S} 1.34 \times 10^{-19} M_{\odot} \frac{\mathrm{GeV}}{\mu},
\end{equation}
where $M_{P l}$ is the Planck's mass, and $\alpha_{B S}$ is a constant numerical parameter which is set to $\alpha_{B S}=1.058$ for static and spherically symmetric metrics \cite{freitas_ultralight_2021}.
Using the definition (\ref{mu}) of $\mu$ we can write $M_{ADM}$ in terms of $m$, $q$ and $\lambda$ as 
\begin{equation}
    M_{A D M}^{\max }=\alpha_{B S} 1.34 \times 10^{-19}M_{\odot} \frac{\sqrt{\lambda}}{\sqrt{2} q m}\mathrm{GeV}.
\end{equation}
In Figures $1$ and $2$, we plot the maximal ADM mass $M_{A D M}^{\max }$ for different values of $\lambda$ and $q$. From Figure $1$, we can observe that for a fixed $\lambda=10^{-50}$ and an effective mass of the scalar field of $m=10^{-24}eV$, the maximal mass $M_{ADM}$ of the Proca star reaches values of astrophysical relevance when $q<10^{-8}$, while for larger values of $q$, the masses are considerably smaller than a solar mass. On the other hand, if we fix $q=10^{-15}$ now, from Figure $2$ we can observe that $M_{ADM}$ increases for large values of $\lambda$. However, due to the constraints of Ly$-$alpha observations adjusted for self-interacting SFDM studied in \cite{li_cosmological_2014}, we know that for ultra-light effective masses of self-interacting scalar fields, the self-interaction parameter $\lambda$ must be very small (but different from zero). In Figure $2$, we consider values of $\lambda \sim 10^{-51}$ that, although not as small as predicted in \cite{li_cosmological_2014}, can be considered within the physical range of SFDM.
Therefore, we can conclude that, as expected, the $\lambda$, $m$, and $q$ parameters of the SFDM field strongly influence the mass of the associated dark photons, and thus also affect the masses of the Proca stars obtained from this system.

\begin{figure}[h]\label{MADMq}
  \centering
  \includegraphics[width=\linewidth]{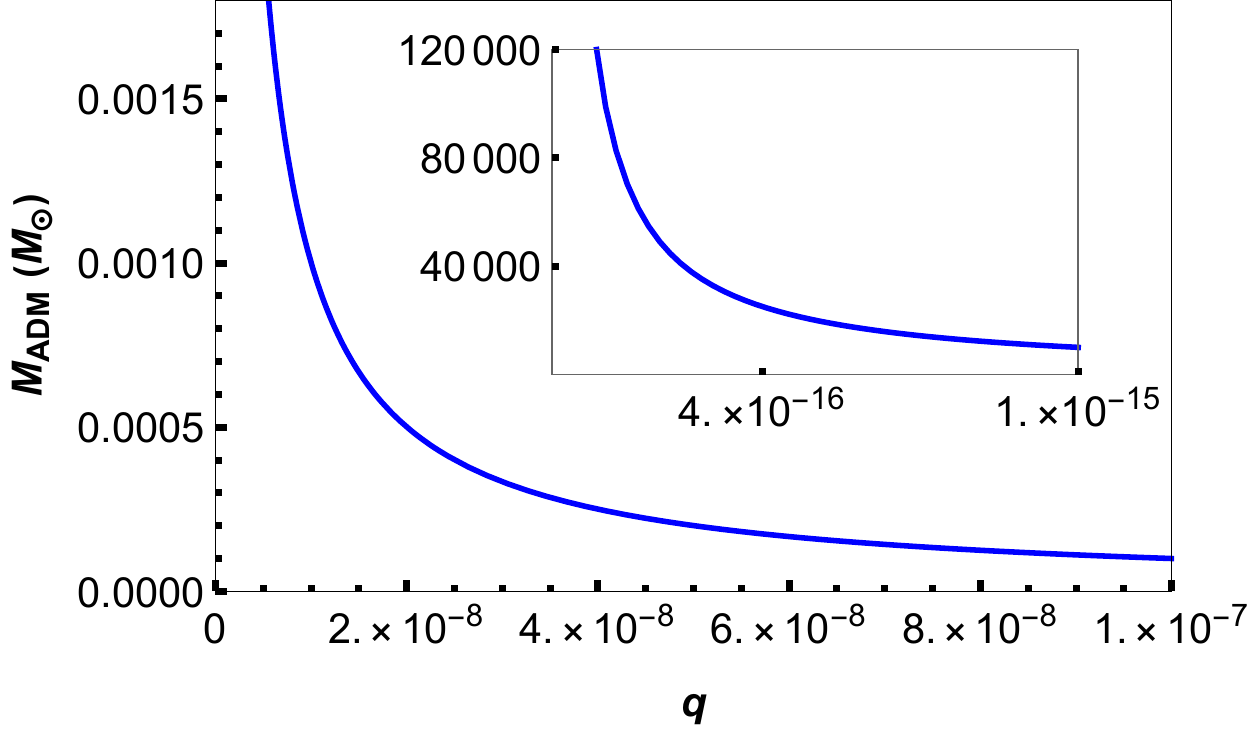}
  \caption{Maximal ADM mass for different values of the gauge charge $q$ with $\lambda=10^{-50}$ and $m=10^{-24}eV$.}
  \label{MQ}
\end{figure}

\begin{figure}[t]\label{MADMlambda}
  \centering
  \includegraphics[width=\linewidth]{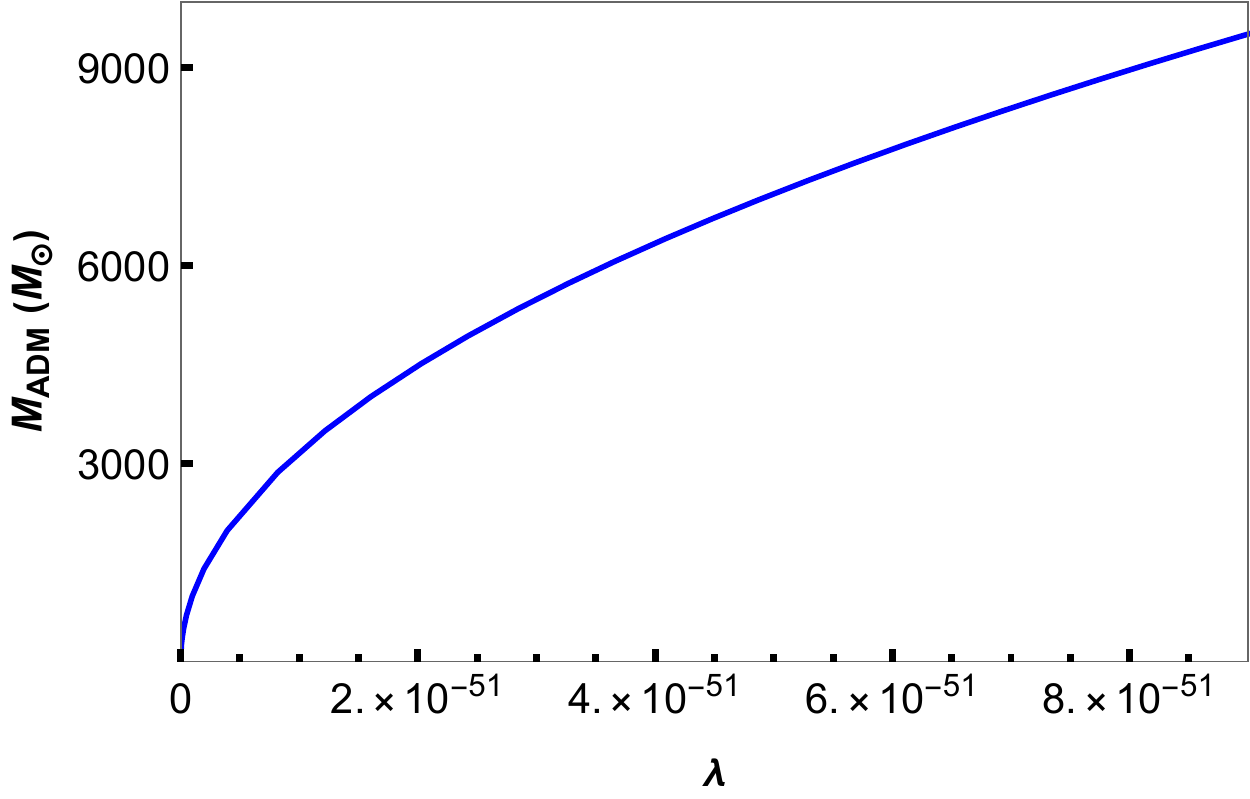}
  \caption{Maximal ADM mass for different values of the self-interaction parameter $\lambda$ of the SFDM field with $q=10^{-15}$ and $m=10^{-24}eV$.}
  \label{MLAMB}
\end{figure}

%% file: sections/PSDP.tex
\section{Proca Stars and Dark Photon} \label{sec:PSDP}
We now consider the case where the electromagnetic field of the SM is not negligible. In this case, the nature of the DP and its interaction with the SM photon becomes manifest, so we consider the Lagrangian (\ref{LDP}), which from now on we call $\mathcal{L}_{DP}$, so that
\begin{equation}
    \begin{aligned}
\mathcal{L}_{DP}= & -\frac{1}{4} F_{\mu \nu} F^{* \mu \nu}-\frac{1}{4} B_{\mu \nu} B^{* \mu \nu} \\
& -U\left(B^{2}\right)-\frac{\delta^{2}}{2} R_{e}\left\{F_{\mu \nu} B^{\mu \nu}\right\},
\end{aligned}
\end{equation}
where $U\left(B^{2}\right)$ is defined the same as in the previous section. Following the same analysis as in the case of Proca, we find that the action for the system is given by $S_{DP}=\int\left[\frac{1}{16 \pi G} R+\mathcal{L}_{DP}\right] \sqrt{-g} d^{4} x$. The variation of this action with respect to the metric leads us to the following Einstein equations
\begin{equation}
    \begin{aligned}
G_{\mu \nu} & =8 \pi G\left(g_{\mu \nu} \mathcal{L}_{D P}-F_{\alpha(\mu} F_{\nu)}^{* \alpha}-B_{\alpha(\mu} B_{\nu)}^{* \alpha}\right. \\
& \left.+2 \hat{U} B_{(\mu} B_{\nu)}^*-\delta^2\left[F_{\alpha(\mu} B_{\nu)}^\alpha+F_{\alpha(\mu}^* B_{\nu)}^{* \alpha}\right]\right).
\end{aligned}
\end{equation}
While varying $A_{\mu}$ and $B_{\mu}$ we obtain the following equations of motion for the fields. Varying $A_{\mu}$ we have
\begin{equation}
    \nabla_{\mu}\left(F^{\mu \nu}+\delta^{2} B^{* \mu \nu}\right)=0,
\end{equation}
and varying $B_{\mu}$ we have
\begin{equation}
    \nabla_{\mu}\left(B^{\mu \nu}+\delta^{2} F^{* \mu \nu}\right)=2 \hat{U} B^{\nu}.
\end{equation}
From these field equations we observe that for the case where the kinetic mixing parameter is $\delta^{2}$$=$$0$, the Maxwell and Proca electrodynamics are recovered for the fields $A_{\mu}$ and $B_{\mu}$ respectively, which was to be expected. To find solutions to the system, we again consider a static and spherically symmetric metric like the one proposed in (\ref{metric}) and, inspired by Proca's ansatz \cite{brito_proca_2016}, we propose that $A_{\mu}$ and $B_ {\mu}$ are of the following form
\begin{equation}\label{B}
    B=e^{-i \omega t}[f(r) d t+i g(r) d r],
\end{equation}
\begin{equation}\label{A}
     A=e^{i \omega t}[h(r) d t-i j(r) d r],
\end{equation}
where, as in the case of Proca, $f(r)$, $g(r)$, $h(r)$ and $j(r)$ are real functions that only depend on the radial coordinate, while $\omega$ is a frequency parameter. Substituting these forms of $A_{\mu}$, $B_{\mu}$ and the metric (\ref{metric}) we obtain that the non-zero Einstein equations are
\begin{equation}\label{1}
    \begin{aligned}
m^{\prime} & =4 \pi G r^{2}\left\{\mu^{2}\left[\frac{N g^{2}}{2}+\frac{f^{2}}{2 N \sigma^{2}}\right]+\frac{1}{2 \sigma^{2}}\left[\left(\omega g-f^{\prime}\right)^{2}\right.\right. \\
& \left.\left.+2 \delta^{2}\left(w g-f^{\prime}\right)\left(\omega j-h^{\prime}\right)+\left(w_{j}-h^{\prime}\right)^{2}\right]\right\},
\end{aligned}
\end{equation}

\begin{equation}\label{2}
    \frac{\sigma^{\prime}}{\sigma}=4 \pi G r \mu^{2}\left[g^{2}+\frac{f^{2}}{N^{2} \sigma^{2}}\right].
\end{equation}
The field equations resulting from varying $B_{\mu}$ are
\begin{equation}\label{3}
    \frac{d}{d r}\left\{\frac{r^{2}}{\sigma}\left[\left(f^{\prime}-\omega g\right)+\delta^{2}\left(h^{\prime}-\omega j\right)\right]\right\}=\frac{r^{2} \mu^{2} f}{\sigma N},
\end{equation}

\begin{equation}\label{4}
    \left(\omega g-f^{\prime}\right)+\delta^{2}\left(\omega j-h^{\prime}\right)=\frac{\mu^{2} \sigma^{2} g N}{\omega},
\end{equation}
and the field equations resulting from varying $A_{\mu}$ are
\begin{equation}\label{5}
    \frac{d}{d r}\left\{\frac{r^{2}}{\sigma}\left[\left(h^{\prime}-\omega j\right)+\delta^{2}\left(f^{\prime}-\omega g\right)\right]\right\}=0,
\end{equation}
\begin{equation}\label{6}
    h^{\prime}-\omega j=-\delta^{2}\left(f^{\prime}-\omega g\right).
\end{equation}
Thus, in principle we have a system of $6$ differential equations with $6$ functions to determine. Note, however, that the penultimate equation (\ref{5}) holds trivially because of the last (\ref{6}). Furthermore we can see that by substituting the equation (\ref{6}) into the first $4$ equations (\ref{1})$-$(\ref{4}), we can eliminate the functions $j(r)$ and $h(r)$ from these equations. Then the system is reduced to $5$ equations with $6$ functions to determine. Now Einstein's equations can be written as
\begin{equation}\label{dp1}
    m^{\prime}=4 \pi G r^{2}\left[\frac{(1-\delta^4)\left( f^{\prime}-\omega g\right)^{2}}{2 \sigma^{2}}+\frac{\mu^{2}}{2} \left(g^{2} N+\frac{f^{2}}{N \sigma^{2}}\right)\right],
\end{equation}
\begin{equation}\label{dp2}
    \frac{\sigma^{\prime}}{\sigma}=4 \pi G r \mu^{2}\left(g^{2}+\frac{f^{2}}{N^{2} \sigma^{2}}\right),
\end{equation}
and the field equations in this case reduce to

\begin{equation}\label{dp3}
    \frac{d}{d r}\left\{\frac{r^{2}\left[f^{\prime}-\omega g\right]}{\sigma}\right\}=\frac{\mu^{2} r^{2} f}{\sigma N (1-\delta^4)},
\end{equation}
\begin{equation}\label{dp4}
    \omega g-f^{\prime}=\frac{\mu^{2} \sigma^{2} N g}{\omega (1-\delta^4)},
\end{equation}

\begin{equation}\label{dp5}
    h^{\prime}-\omega j=-\delta^{2}\left(f^{\prime}-\omega g\right).
\end{equation}
    
We can see that the first $4$ equations (\ref{dp1})$-$(\ref{dp4}) of this new system simply correspond to a system of equations of the type Einstein-Proca as obtained in the Proca stars section above, except for a constant term $(1-\delta^4)$ that multiplies the equations. The constraints for $\delta$ are $\delta^{2} \neq 1 ; \delta^{2} \leqslant 10^{-6}$ \cite{caputo_dark_2021} (because very weak interactions are requested between both photons), therefore $\left(1-\delta^{4}\right) \neq 0$ and the system is well defined. While the functions $h(r)$ and $j(r)$ that describe the field of the SM photon are defined in terms of those of the dark photon $f(r)$ and $g(r)$ through the equation (\ref{dp5}). To close the system we can consider a gauge fixing for $A_{\mu}$, such as the Lorentz condition $\nabla_{\mu} A^{\mu}=0$ (which is a requirement and not a choice for the case of $B_{\mu}$ \cite{brito_proca_2016}). In this case, considering the ansatz (\ref{A}) and the metric (\ref{metric}), the Lorentz condition for $A_{\mu}$ can be written as 
\begin{equation}\label{dp6}
   \frac{d}{d r}\left\{\frac{\sigma j r^2 N}{\omega}\right\}=\frac{-r^2 h}{\sigma N}.
\end{equation}

In order to find the solutions for the functions that describe the system, we must simultaneously solve the equations (\ref{dp1})$-$(\ref{dp6}). 
In the next section we solve the system numerically for the non-asymptotic regions, and we study graphically how the solutions for each function are modified. We are particularly interested in the functions $j(r)$ and $h(r)$ that describe the photon of the SM, since, in principle, it is from these functions that observations could be obtained.

%% file: sections/Results.tex
\section{Results} \label{sec:Results}
To preserve the regularity of the system of equations at the origin, it is necessary to consider series expansions for the functions around $r = 0$, so that \cite{brito_proca_2016, rosa_shadows_2022}
\begin{equation}
    \begin{aligned}
 f(r)&=f_0+\mathcal{O}\left(r^2\right), \\
 g(r)&=\mathcal{O}(r), \\
 m(r)&=\mathcal{O}\left(r^3\right), \\
 \sigma(r)&=\sigma_0+\mathcal{O}\left(r^2\right), \\
 h(r)&=h_0+\mathcal{O}\left(r^2\right), \\
 j(r)&=\mathcal{O}(r),
\end{aligned}
\end{equation}
where $\sigma(0)=\sigma_0$, $f(0)=f_0$ and $h(0)=h_0$ are the boundary conditions at the origin.
Additionally, to preserve the asymptotic flatness, it is necessary that the mass functions $f$ and $g$ of the dark photon decay exponentially at infinity, therefore
\begin{equation}
    f(\infty)=g(\infty)=0.
\end{equation}
While the functions $N$ and $\sigma$ of the metric must approach $1$ at infinity. Therefore we impose
\begin{equation}
    m(\infty)=M, \quad \sigma(\infty)=1,
\end{equation}
where $M$ is the total mass of the star. Finally, for the electromagnetic field of the SM photon, we can consider a constant electric potential and a magnetic potential equal to zero at infinity
\begin{equation}
    h(\infty)=\text { constant }, \quad j(\infty)=0.
\end{equation}
The numerical results are calculated using the rescaled quantities $r \rightarrow \mu r$, $m \rightarrow \mu m$, $\omega \rightarrow \omega / \mu$ and the rescaled potentials $f \rightarrow \sqrt{4 \pi G} f$, $g \rightarrow \sqrt{4 \pi G} g$, $h \rightarrow \sqrt{4 \pi G} h$, and $j \rightarrow \sqrt{4 \pi G} j$.
To solve the system we use $"\text{NDSolve}"$ command in $Mathematica$ and evaluate the initial conditions at $r=10^{-6}$. We can vary $f_0$ in analogy to a shooting method and set $\sigma_0$ and $M$ numerically. 

In Figures $3$ and $4$, we use approximately the same values for the parameters as in Figure 2 in \cite{brito_proca_2016} to compare how the Proca case solutions differ due to the presence of the SM photon. In both figures, we take $ \delta^{2} \sim10^{-14}$ based in the Ly$-$alpha observations adjusted for dark photons in reference \cite{bolton_comparison_2022}. As we can observe, the functions $m(r)$ and $\sigma(r)$ of the metric and the functions $f(r)$ and $g(r)$ of the dark photon vary very little from the case where $\delta^{2}=0$ (the pure Proca case presented in \cite{brito_proca_2016}).
This is to be expected due to the $\delta^4$ factor in the differential equations. In fact, at a given numerical radius, $R$, the variations of the numerical solutions for the functions $m(r)$ and $\sigma(r)$ when $\delta^{2}=0$ and when $ \delta^{2} \sim10^{-14}$ are of the order of $m(\delta^2=0)-m(\delta^2=10^{-14}) \sim 10^{-7}$. Variations in these functions become noticeable until the kinetic mixing parameter $\delta^2$ takes on values so large that they are outside of the physical ultralight DP range. In Figure $5$, for example, we take $\delta^2>10^{-3}$; in these cases, the changes in the mass function $m(r)$ for the star become considerable, and therefore changes in the maximal ADM mass could be expected; however, taking values of $\delta$ in the physical range, these differences become increasingly negligible. The solutions for the functions $g(r)$, $f(r)$, $j(r)$, and $h(r)$ present similar behavior with respect to the variations in $\delta^2$. 

\begin{figure}[b]\label{functions01}
  \centering
  \includegraphics[width=\linewidth]{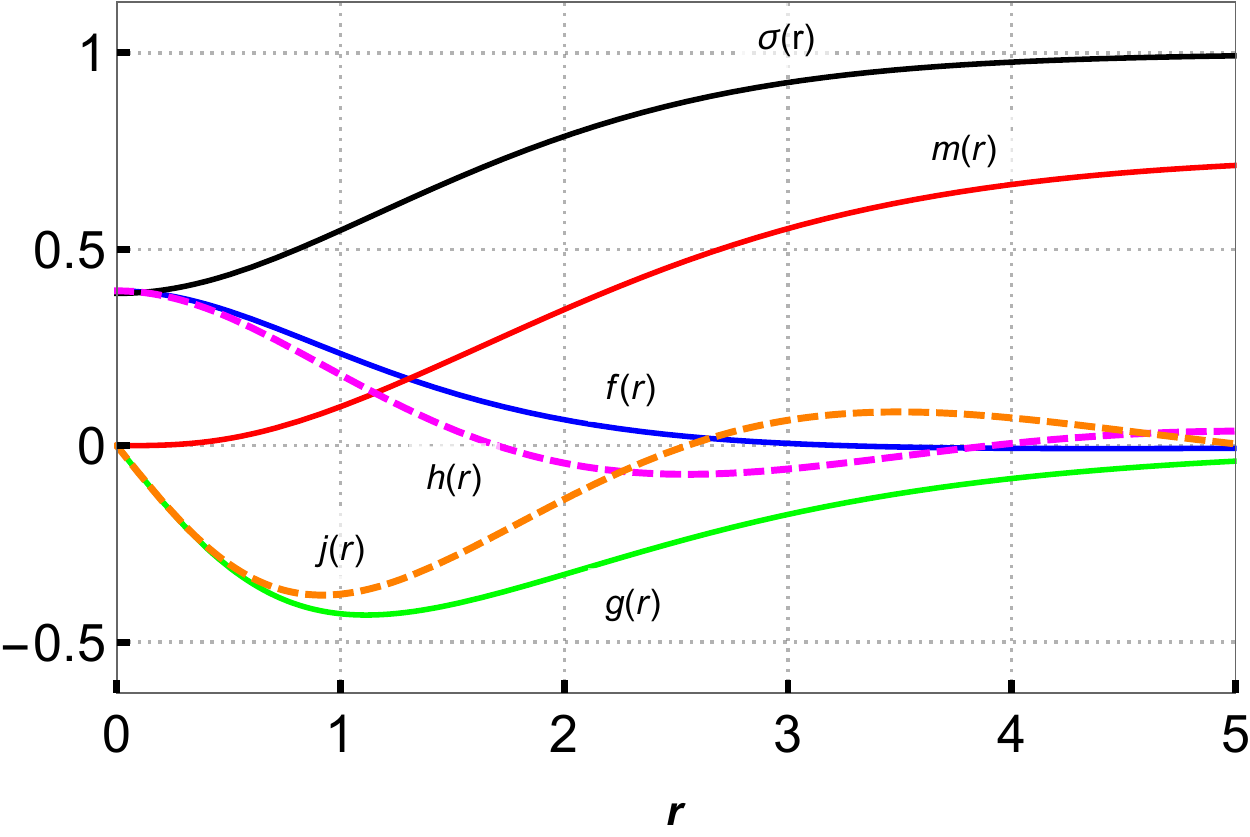}
  \caption{Numerical solutions for the functions $m(r)$ (red), $\sigma(r)$ (black), $f(r)$ (blue), $g(r)$ (green), $h(r)$ (dashed magenta), and $j(r)$ (dashed orange) are obtained with fixed $M=0.745$, $\omega=0.817$, $f_0=h_0=0.394$, and $\delta^2=10^{-14}$.}
  \label{FUNC1}
\end{figure}

\begin{figure}[t]\label{functions02}
  \centering
  \includegraphics[width=\linewidth]{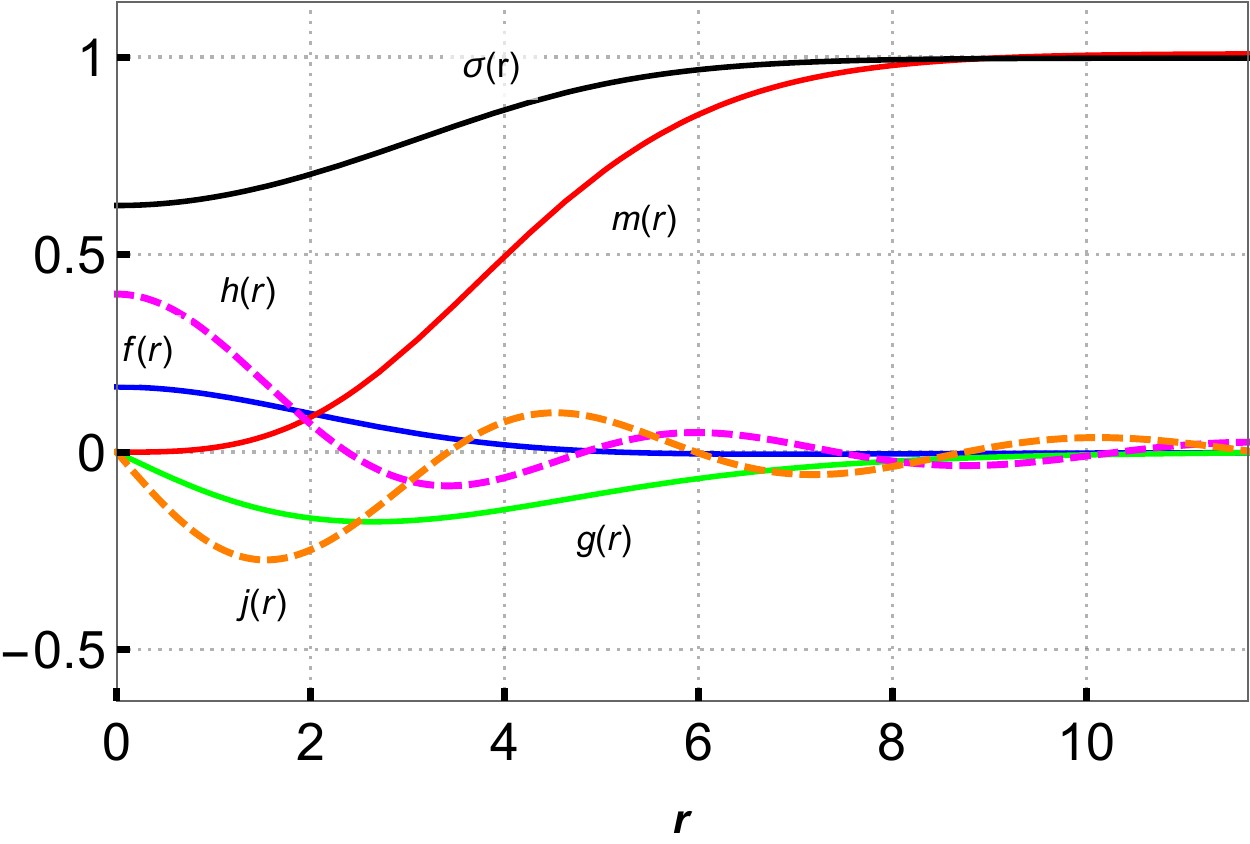}
  \caption{Numerical solutions for the functions $m(r)$ (red), $\sigma(r)$ (black), $f(r)$ (blue), $g(r)$ (green), $h(r)$ (dashed magenta), and $j(r)$ (dashed orange) are obtained with fixed $M=1.016$, $\omega=0.839$, $f_0=0.165$, $h_0=0.4$ and $\delta^2=10^{-14}$.}
  \label{FUNC2}
\end{figure}

\begin{figure}[b]\label{mf0}
  \centering
  \includegraphics[width=\linewidth]{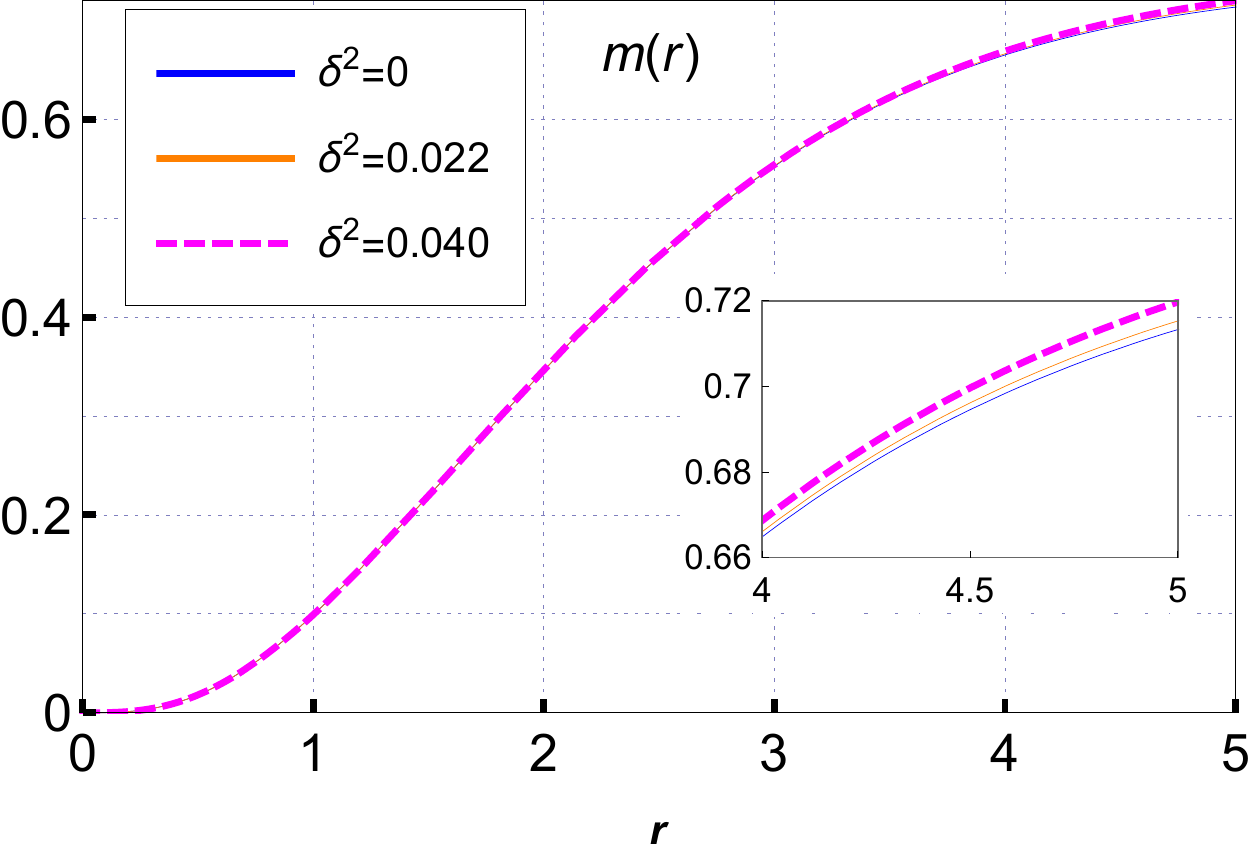}
  \caption{Numerical solutions for the function $m(r)$ are obtained for different values of $\delta^{2}$ with fixed $\omega=0.817$, and $f_0=h_0=0.394$.}
  \label{M0GR}
\end{figure}

\begin{figure}[t]\label{h0}
  \centering
  \includegraphics[width=\linewidth]{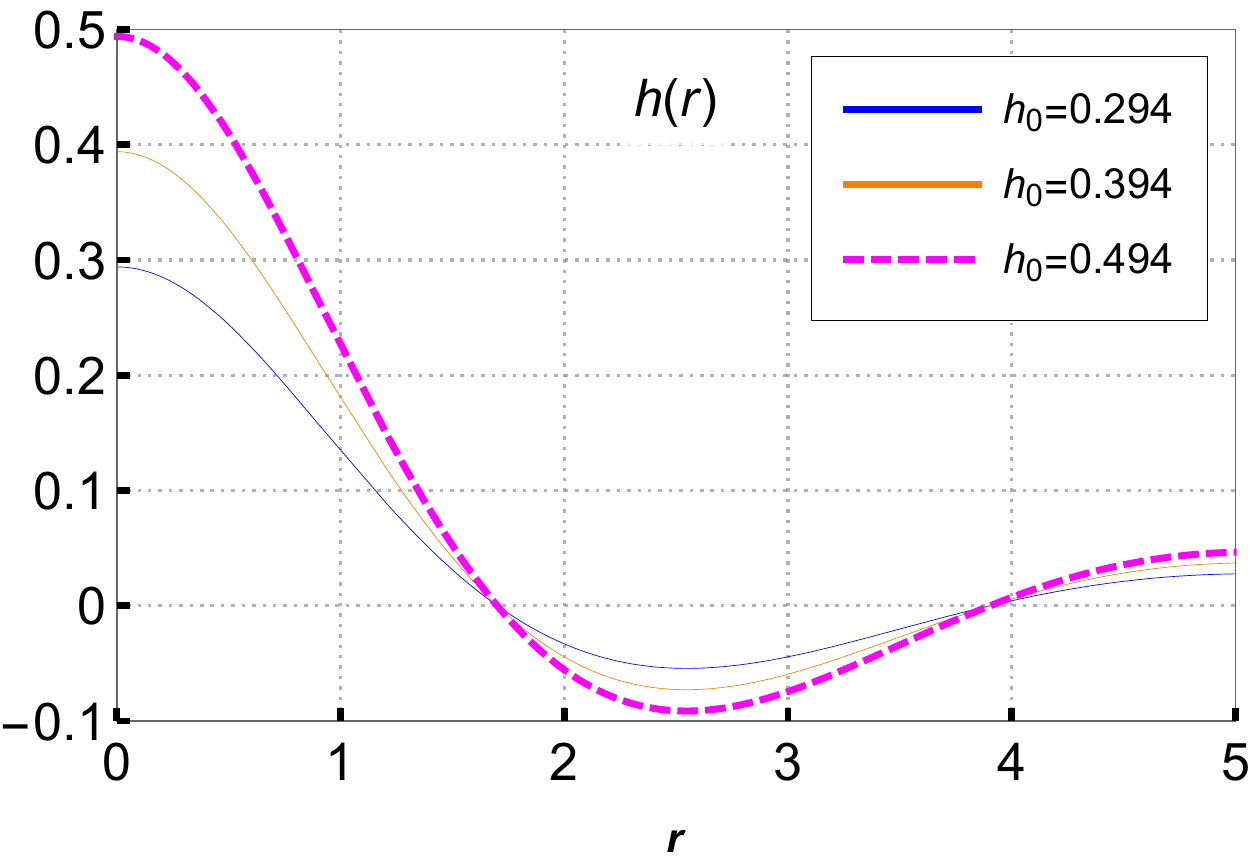}
  \caption{Numerical solutions for the function $h(r)$ are obtained for different values of $h_0$ with fixed $M=0.745$, $\omega=0.817$, $f_0=0.394$, and $\delta^{2}=10^{-14}$.}
  \label{H0GR}
\end{figure}

On the other hand, due to the way the functions are coupled in the system of differential equations, we know that the initial values $f_0$ and $\sigma_0$ strongly determine the behavior of the metric and DP functions, while the initial value $h_0$ of the SM photon has a weak influence on these functions. In fact, by varying $h_0$ as shown in Figure $6$, the difference between the numerical solutions for the functions $f(r)$, $g(r)$, $\sigma(r)$, and $m(r)$ at a given numerical radius $R$ is of the order of $\sim 10^{-6}$. This suggests that the SM photon has a relatively small effect on the behavior of the other functions in the system and that it may be possible to neglect it in certain cases. However, the choice of $h_0$ strongly influences the functions of the SM photon $j(r)$ and $h(r)$. Therefore, we can vary $h_0$ to fix the values of $j(r)$ and $h(r)$ at infinity without significantly modifying the metric and DP functions. This would allow us, in principle, to vary $h_0$ to adjust the observables of the electromagnetic field of the SM photon associated with a Proca-like star.

%% file: sections/Observables.tex
\section{Observables} \label{sec:Observables}
Due to the form of the ansatz (\ref{A}) for $A_{\mu}$, the solutions are associated with a system with zero magnetic field $\bar{B}=0$ and an electric field $\bar{E}$ given in terms of the electromagnetic tensor $F_{\mu \nu}$ as follows 
\begin{equation}
    E_i=F_{i 0},
\end{equation}
where $i=\{r, \theta, \phi\}$, so that
\begin{equation}
\bar{E}=\left[h^{\prime}(r)-\omega j(r)\right] e^{i \omega t} \hat{r},    
\end{equation}
where $\hat{r}$ is the unit vector in the radial direction. Therefore, the physical electric field is given by
\begin{equation}
    \operatorname{Re}\{\bar{E}\}=\left[h^{\prime}(r)-w j(r)\right] \cos (w t) \hat{r},
\end{equation}
and from now on we change the notation for the electric field to $\bar{E} \rightarrow \operatorname{Re}\{\bar{E}\}$. From here, we can calculate the expected value of the volumetric energy density associated with this electric field, given by
\begin{equation}
    \left\langle\frac{1}{2} E^2\right\rangle=\frac{1}{4}\left[h^{\prime}(r)-w j(r)\right]^2.
\end{equation}
In order to use the numerical solutions calculated in the previous section, it is necessary to return to the non-scaled functions. Therefore, in natural units, the electric field can be written in terms of the numerical solutions as follows
\begin{equation}\label{E2num}
    \left\langle\frac{1}{2} E^2\right\rangle=\frac{\mu^2}{16 \pi G}\left[h_{(\text {num })}^{\prime}-w_{(\text {num })} j_{(\text {num })}\right]^2.
\end{equation}

Finally, in order to avoid numerical noise, we can use equation (\ref{dp5}) to rewrite $h(r)$ and $j(r)$ in terms of $f(r)$ and $g(r)$, and directly introduce the $\delta^4$ factor in the previous equation, so that
\begin{equation}\label{energySI}
    \left\langle\frac{1}{2} E^2\right\rangle=\frac{\mu^2 \delta^4}{16 \pi G}\left[f_{(\text {num })}^{\prime}-w_{(\text {num })} g_{(\text {num })}\right]^2.
\end{equation}

\begin{figure}[t]\label{E2}
  \centering
  \includegraphics[width=\linewidth]{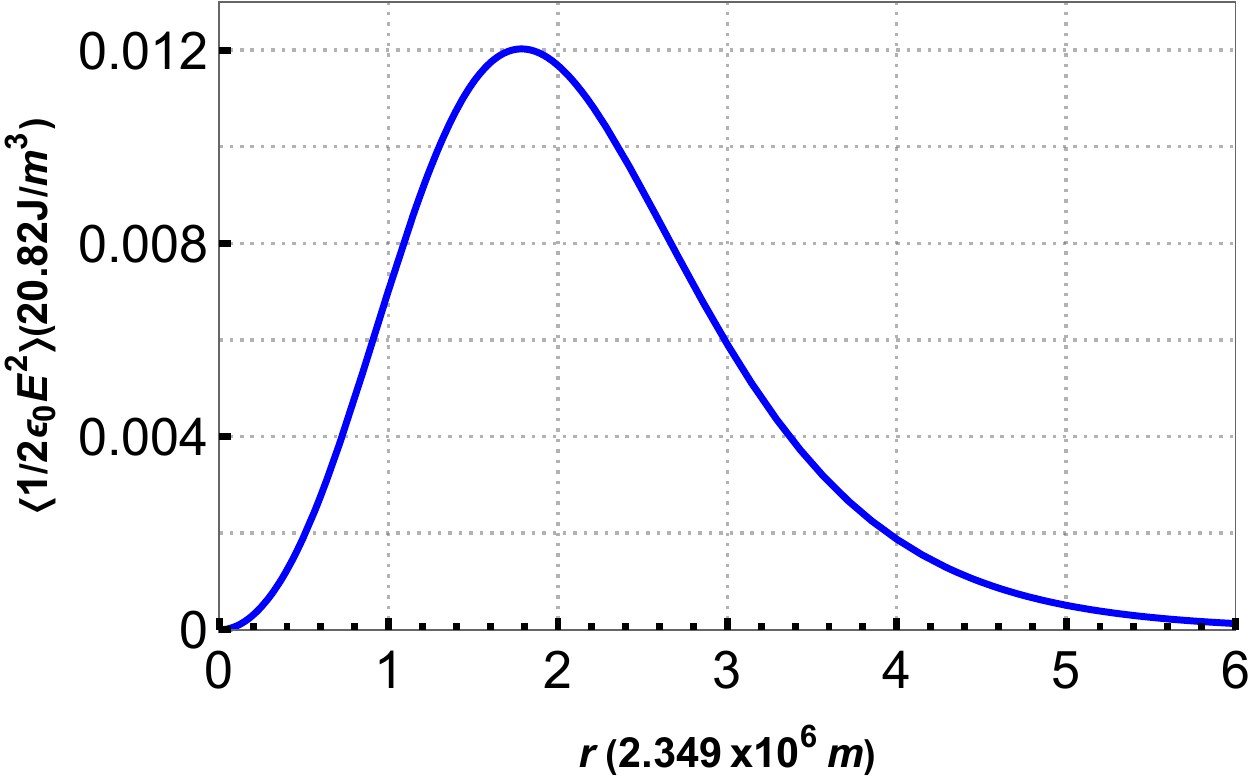}
  \caption{The expected value for the volumetric energy density of the electric field $E$ is obtained for a solution with $\mu=8.4\times10^{-14}$ and $\delta^{2}=4.6\times10^{-15}$.}
  \label{E2SI}
\end{figure}

In Figure 7, we present the density plot for numerical solutions with fixed values of $M=0.745$, $\omega=0.817$, $f_0=h_0=0.394$, and $\delta^{2}=4.6\times10^{-15}$. We then return to the non-scaled functions by using a value of $\mu=8.4\times10^{-14}eV$ (motivated by the simulations made in \cite{bolton_comparison_2022} for DP) and express the results in SI units. For these specific parameter values, we obtain a maximum for the volumetric energy density at approximately $0.24 \mathrm{~J} / \mathrm{m}^3$. As we can observe, due equation (\ref{energySI}), the expected value of the volumetric energy density of the electric field strongly depends on the chosen value of $\delta^2$. This energy decays to zero at infinity, as expected. In principle, it is this electric field and this density that we can use to adjust observations in the regions where Proca stars associated with dark photons are expected to be found. We hope to use this observable in future works to adjust real observational data.

%% file: sections/section03.tex
\section{Conclusions} \label{sec:conclusions}
    From this proposed model where SFDM possesses a gauge symmetry $U(1)$ whose gauge charge is associated with the dark photon, it is possible to obtain the description of scalar boson stars, both charged and uncharged. Moreover, by considering a spontaneous symmetry breaking in which the dark photon acquires mass, it is possible to obtain Proca-type stars composed of these dark photons.

   In the case where the mixing between the dark photon and the SM photon is negligible, the result is the usual mini-Proca stars described in \cite{brito_proca_2016}, with the only difference being that the mass of the DP is in terms of the effective mass, charge, and self-coupling parameter of the dark matter scalar field. This causes the maximal mass of these Proca stars to be modified in terms of the scalar field parameters.
 
 On the other hand, the presence of the kinetic mixing term between the DP and the SM photon has an influence on the solutions for Proca-type stars. Specifically, the larger the kinetic mixing parameter $\delta^2$, the further the solutions deviate from the pure Proca case. However, when $\delta^2$ takes physical values under the ultralight DP context, the differences in the numerical solutions become increasingly negligible.

Finally, we used the numerical results to calculate the electric field and volumetric energy density associated with the SM electromagnetic field for the case of Proca stars formed by dark photons when the mixing between both photons is not negligible. This is an important result that can help investigate physical observables for these types of stars, in addition to the gravitational effects that are already known.

%% file: sections/acknowledgements.tex
\section*{Acknowledgements} \label{sec:acknowledgements}
     This work was partially supported by CONACyT M\'exico under grants  A1-S-8742, 304001, 376127, 240512,  FORDECYT-PRONACES grant No. 490769 and I0101/131/07 C-234/07 of the Instituto Avanzado de Cosmolog\'ia (IAC) collaboration (http://www.iac.edu.mx/).

%% file: main.bbl
\begin{thebibliography}{87}%
\makeatletter
\providecommand \@ifxundefined [1]{%
 \@ifx{#1\undefined}
}%
\providecommand \@ifnum [1]{%
 \ifnum #1\expandafter \@firstoftwo
 \else \expandafter \@secondoftwo
 \fi
}%
\providecommand \@ifx [1]{%
 \ifx #1\expandafter \@firstoftwo
 \else \expandafter \@secondoftwo
 \fi
}%
\providecommand \natexlab [1]{#1}%
\providecommand \enquote  [1]{``#1''}%
\providecommand \bibnamefont  [1]{#1}%
\providecommand \bibfnamefont [1]{#1}%
\providecommand \citenamefont [1]{#1}%
\providecommand \href@noop [0]{\@secondoftwo}%
\providecommand \href [0]{\begingroup \@sanitize@url \@href}%
\providecommand \@href[1]{\@@startlink{#1}\@@href}%
\providecommand \@@href[1]{\endgroup#1\@@endlink}%
\providecommand \@sanitize@url [0]{\catcode `\\12\catcode `\$12\catcode
  `\&12\catcode `\#12\catcode `\^12\catcode `\_12\catcode `\%12\relax}%
\providecommand \@@startlink[1]{}%
\providecommand \@@endlink[0]{}%
\providecommand \url  [0]{\begingroup\@sanitize@url \@url }%
\providecommand \@url [1]{\endgroup\@href {#1}{\urlprefix }}%
\providecommand \urlprefix  [0]{URL }%
\providecommand \Eprint [0]{\href }%
\providecommand \doibase [0]{https://doi.org/}%
\providecommand \selectlanguage [0]{\@gobble}%
\providecommand \bibinfo  [0]{\@secondoftwo}%
\providecommand \bibfield  [0]{\@secondoftwo}%
\providecommand \translation [1]{[#1]}%
\providecommand \BibitemOpen [0]{}%
\providecommand \bibitemStop [0]{}%
\providecommand \bibitemNoStop [0]{.\EOS\space}%
\providecommand \EOS [0]{\spacefactor3000\relax}%
\providecommand \BibitemShut  [1]{\csname bibitem#1\endcsname}%
\let\auto@bib@innerbib\@empty
\bibitem [{\citenamefont {Ji}\ and\ \citenamefont
  {Sin}(1994{\natexlab{a}})}]{ji_late-time_1994}%
  \BibitemOpen
  \bibfield  {author} {\bibinfo {author} {\bibfnamefont {S.~U.}\ \bibnamefont
  {Ji}}\ and\ \bibinfo {author} {\bibfnamefont {S.~J.}\ \bibnamefont {Sin}},\
  }\bibfield  {title} {{\selectlanguage {english}\bibinfo {title} {Late-time
  phase transition and the galactic halo as a {Bose} liquid. {II}. {The} effect
  of visible matter}},\ }\href {https://doi.org/10.1103/PhysRevD.50.3655}
  {\bibfield  {journal} {\bibinfo  {journal} {Phys. Rev. D}\ }\textbf {\bibinfo
  {volume} {50}},\ \bibinfo {pages} {3655} (\bibinfo {year}
  {1994}{\natexlab{a}})}\BibitemShut {NoStop}%
\bibitem [{\citenamefont {Lee}\ and\ \citenamefont
  {Koh}(1996)}]{lee_galactic_1996}%
  \BibitemOpen
  \bibfield  {author} {\bibinfo {author} {\bibfnamefont {J.-w.}\ \bibnamefont
  {Lee}}\ and\ \bibinfo {author} {\bibfnamefont {I.-g.}\ \bibnamefont {Koh}},\
  }\bibfield  {title} {{\selectlanguage {english}\bibinfo {title} {Galactic
  {Halos} {As} {Boson} {Stars}}},\ }\href
  {https://doi.org/10.1103/PhysRevD.53.2236} {\bibfield  {journal} {\bibinfo
  {journal} {Phys. Rev. D}\ }\textbf {\bibinfo {volume} {53}},\ \bibinfo
  {pages} {2236} (\bibinfo {year} {1996})},\ \bibinfo {note}
  {arXiv:hep-ph/9507385}\BibitemShut {NoStop}%
\bibitem [{\citenamefont {Guzman}\ and\ \citenamefont
  {Matos}(2000)}]{guzman_scalar_2000}%
  \BibitemOpen
  \bibfield  {author} {\bibinfo {author} {\bibfnamefont {F.~S.}\ \bibnamefont
  {Guzman}}\ and\ \bibinfo {author} {\bibfnamefont {T.}~\bibnamefont {Matos}},\
  }\bibfield  {title} {{\selectlanguage {english}\bibinfo {title} {Scalar
  {Fields} as {Dark} {Matter} in {Spiral} {Galaxies}}},\ }\href
  {https://doi.org/10.1088/0264-9381/17/1/102} {\bibfield  {journal} {\bibinfo
  {journal} {Class. Quantum Grav.}\ }\textbf {\bibinfo {volume} {17}},\
  \bibinfo {pages} {L9} (\bibinfo {year} {2000})},\ \bibinfo {note}
  {arXiv:gr-qc/9810028}\BibitemShut {NoStop}%
\bibitem [{\citenamefont {Maga{\~n}a}\ \emph {et~al.}(2012)\citenamefont
  {Maga{\~n}a}, \citenamefont {Matos}, \citenamefont {Robles},\ and\
  \citenamefont {Su{\'a}rez}}]{magana_brief_2012}%
  \BibitemOpen
  \bibfield  {author} {\bibinfo {author} {\bibfnamefont {J.}~\bibnamefont
  {Maga{\~n}a}}, \bibinfo {author} {\bibfnamefont {T.}~\bibnamefont {Matos}},
  \bibinfo {author} {\bibfnamefont {V.}~\bibnamefont {Robles}},\ and\ \bibinfo
  {author} {\bibfnamefont {A.}~\bibnamefont {Su{\'a}rez}},\ }\bibfield  {title}
  {{\selectlanguage {english}\bibinfo {title} {A brief {Review} of the {Scalar}
  {Field} {Dark} {Matter} model}},\ }\href
  {https://doi.org/10.1088/1742-6596/378/1/012012} {\bibfield  {journal}
  {\bibinfo  {journal} {J. Phys.: Conf. Ser.}\ }\textbf {\bibinfo {volume}
  {378}},\ \bibinfo {pages} {012012} (\bibinfo {year} {2012})},\ \bibinfo
  {note} {arXiv:1201.6107 [astro-ph]}\BibitemShut {NoStop}%
\bibitem [{\citenamefont {Ure{\~n}a-L{\'o}pez}(2019)}]{urena-lopez_brief_2019}%
  \BibitemOpen
  \bibfield  {author} {\bibinfo {author} {\bibfnamefont {L.~A.}\ \bibnamefont
  {Ure{\~n}a-L{\'o}pez}},\ }\bibfield  {title} {{\selectlanguage
  {english}\bibinfo {title} {Brief {Review} on {Scalar} {Field} {Dark} {Matter}
  {Models}}},\ }\href {https://doi.org/10.3389/fspas.2019.00047} {\bibfield
  {journal} {\bibinfo  {journal} {Front. Astron. Space Sci.}\ }\textbf
  {\bibinfo {volume} {6}},\ \bibinfo {pages} {47} (\bibinfo {year}
  {2019})}\BibitemShut {NoStop}%
\bibitem [{\citenamefont {Su{\'a}rez}\ \emph {et~al.}(2014)\citenamefont
  {Su{\'a}rez}, \citenamefont {Robles},\ and\ \citenamefont
  {Matos}}]{suarez_review_2014}%
  \BibitemOpen
  \bibfield  {author} {\bibinfo {author} {\bibfnamefont {A.}~\bibnamefont
  {Su{\'a}rez}}, \bibinfo {author} {\bibfnamefont {V.}~\bibnamefont {Robles}},\
  and\ \bibinfo {author} {\bibfnamefont {T.}~\bibnamefont {Matos}},\ }\bibfield
   {title} {{\selectlanguage {english}\bibinfo {title} {A {Review} on the
  {Scalar} {Field}/ {Bose}-{Einstein} {Condensate} {Dark} {Matter} {Model}}}\
  }(\bibinfo {year} {2014})\ pp.\ \bibinfo {pages} {107--142},\ \bibinfo {note}
  {arXiv:1302.0903 [astro-ph]}\BibitemShut {NoStop}%
\bibitem [{\citenamefont {Rindler-Daller}\ and\ \citenamefont
  {Shapiro}(2014)}]{rindler-daller_complex_2014}%
  \BibitemOpen
  \bibfield  {author} {\bibinfo {author} {\bibfnamefont {T.}~\bibnamefont
  {Rindler-Daller}}\ and\ \bibinfo {author} {\bibfnamefont {P.~R.}\
  \bibnamefont {Shapiro}},\ }\bibfield  {title} {{\selectlanguage
  {english}\bibinfo {title} {Complex scalar field dark matter on galactic
  scales}},\ }\href {https://doi.org/10.1142/S021773231430002X} {\bibfield
  {journal} {\bibinfo  {journal} {Mod. Phys. Lett. A}\ }\textbf {\bibinfo
  {volume} {29}},\ \bibinfo {pages} {1430002} (\bibinfo {year} {2014})},\
  \bibinfo {note} {arXiv:1312.1734 [astro-ph, physics:gr-qc,
  physics:hep-ph]}\BibitemShut {NoStop}%
\bibitem [{\citenamefont {Marsh}(2016)}]{marsh_warmandfuzzy_2016}%
  \BibitemOpen
  \bibfield  {author} {\bibinfo {author} {\bibfnamefont {D.~J.~E.}\
  \bibnamefont {Marsh}},\ }\href {http://arxiv.org/abs/1605.05973}
  {{\selectlanguage {english}\bibinfo {title} {{WarmAndFuzzy}: the halo model
  beyond {CDM}}}} (\bibinfo {year} {2016}),\ \bibinfo {note} {arXiv:1605.05973
  [astro-ph, physics:hep-ph]}\BibitemShut {NoStop}%
\bibitem [{\citenamefont {Niemeyer}(2020)}]{niemeyer_small-scale_2020}%
  \BibitemOpen
  \bibfield  {author} {\bibinfo {author} {\bibfnamefont {J.~C.}\ \bibnamefont
  {Niemeyer}},\ }\bibfield  {title} {{\selectlanguage {english}\bibinfo {title}
  {Small-scale structure of fuzzy and axion-like dark matter}},\ }\href
  {https://doi.org/10.1016/j.ppnp.2020.103787} {\bibfield  {journal} {\bibinfo
  {journal} {Progress in Particle and Nuclear Physics}\ }\textbf {\bibinfo
  {volume} {113}},\ \bibinfo {pages} {103787} (\bibinfo {year} {2020})},\
  \bibinfo {note} {arXiv:1912.07064 [astro-ph]}\BibitemShut {NoStop}%
\bibitem [{\citenamefont {Hu}\ \emph {et~al.}(2000)\citenamefont {Hu},
  \citenamefont {Barkana},\ and\ \citenamefont {Gruzinov}}]{hu_cold_2000}%
  \BibitemOpen
  \bibfield  {author} {\bibinfo {author} {\bibfnamefont {W.}~\bibnamefont
  {Hu}}, \bibinfo {author} {\bibfnamefont {R.}~\bibnamefont {Barkana}},\ and\
  \bibinfo {author} {\bibfnamefont {A.}~\bibnamefont {Gruzinov}},\ }\bibfield
  {title} {{\selectlanguage {english}\bibinfo {title} {Cold and {Fuzzy} {Dark}
  {Matter}}},\ }\href {https://doi.org/10.1103/PhysRevLett.85.1158} {\bibfield
  {journal} {\bibinfo  {journal} {Phys. Rev. Lett.}\ }\textbf {\bibinfo
  {volume} {85}},\ \bibinfo {pages} {1158} (\bibinfo {year} {2000})},\ \bibinfo
  {note} {arXiv:astro-ph/0003365}\BibitemShut {NoStop}%
\bibitem [{\citenamefont {Arbey}\ \emph {et~al.}(2001)\citenamefont {Arbey},
  \citenamefont {Lesgourgues},\ and\ \citenamefont
  {Salati}}]{arbey_quintessential_2001}%
  \BibitemOpen
  \bibfield  {author} {\bibinfo {author} {\bibfnamefont {A.}~\bibnamefont
  {Arbey}}, \bibinfo {author} {\bibfnamefont {J.}~\bibnamefont {Lesgourgues}},\
  and\ \bibinfo {author} {\bibfnamefont {P.}~\bibnamefont {Salati}},\
  }\bibfield  {title} {{\selectlanguage {english}\bibinfo {title}
  {Quintessential {Haloes} around {Galaxies}}},\ }\href
  {https://doi.org/10.1103/PhysRevD.64.123528} {\bibfield  {journal} {\bibinfo
  {journal} {Phys. Rev. D}\ }\textbf {\bibinfo {volume} {64}},\ \bibinfo
  {pages} {123528} (\bibinfo {year} {2001})},\ \bibinfo {note}
  {arXiv:astro-ph/0105564}\BibitemShut {NoStop}%
\bibitem [{\citenamefont {Bray}(2010)}]{bray_wave_nodate}%
  \BibitemOpen
  \bibfield  {author} {\bibinfo {author} {\bibfnamefont {H.~L.}\ \bibnamefont
  {Bray}},\ }\bibfield  {title} {{\selectlanguage {english}\bibinfo {title} {On
  dark matter, spiral galaxies, and the axioms of general relativity}},\
  }\href@noop {} {\bibfield  {journal} {\bibinfo  {journal} {physics.gen-ph}\ }
  (\bibinfo {year} {2010})},\ \Eprint {https://arxiv.org/abs/1004.4016}
  {1004.4016} \BibitemShut {NoStop}%
\bibitem [{\citenamefont {Ferreira}(2021)}]{ferreira_ultra-light_2021}%
  \BibitemOpen
  \bibfield  {author} {\bibinfo {author} {\bibfnamefont {E.~G.~M.}\
  \bibnamefont {Ferreira}},\ }\bibfield  {title} {{\selectlanguage
  {english}\bibinfo {title} {Ultra-{Light} {Dark} {Matter}}},\ }\href
  {https://doi.org/10.1007/s00159-021-00135-6} {\bibfield  {journal} {\bibinfo
  {journal} {Astron Astrophys Rev}\ }\textbf {\bibinfo {volume} {29}},\
  \bibinfo {pages} {7} (\bibinfo {year} {2021})},\ \bibinfo {note}
  {arXiv:2005.03254 [astro-ph, physics:cond-mat, physics:gr-qc,
  physics:hep-th]}\BibitemShut {NoStop}%
\bibitem [{\citenamefont {Matos}(2022)}]{matos_quantum_2022}%
  \BibitemOpen
  \bibfield  {author} {\bibinfo {author} {\bibfnamefont {T.}~\bibnamefont
  {Matos}},\ }\bibfield  {title} {{\selectlanguage {english}\bibinfo {title}
  {The quantum character of the {Scalar} {Field} {Dark} {Matter}}},\ }\href
  {https://doi.org/10.1093/mnras/stac3079} {\bibfield  {journal} {\bibinfo
  {journal} {Monthly Notices of the Royal Astronomical Society}\ }\textbf
  {\bibinfo {volume} {517}},\ \bibinfo {pages} {5247} (\bibinfo {year}
  {2022})},\ \bibinfo {note} {arXiv:2211.02025 [astro-ph]}\BibitemShut
  {NoStop}%
\bibitem [{\citenamefont {Sol{\'i}s-L{\'o}pez}\ \emph
  {et~al.}(2021)\citenamefont {Sol{\'i}s-L{\'o}pez}, \citenamefont
  {Guzm{\'a}n}, \citenamefont {Matos}, \citenamefont {Robles},\ and\
  \citenamefont {Ure{\~n}a-L{\'o}pez}}]{solis-lopez_scalar_2021}%
  \BibitemOpen
  \bibfield  {author} {\bibinfo {author} {\bibfnamefont {J.}~\bibnamefont
  {Sol{\'i}s-L{\'o}pez}}, \bibinfo {author} {\bibfnamefont {F.~S.}\
  \bibnamefont {Guzm{\'a}n}}, \bibinfo {author} {\bibfnamefont
  {T.}~\bibnamefont {Matos}}, \bibinfo {author} {\bibfnamefont {V.~H.}\
  \bibnamefont {Robles}},\ and\ \bibinfo {author} {\bibfnamefont {L.~A.}\
  \bibnamefont {Ure{\~n}a-L{\'o}pez}},\ }\bibfield  {title} {{\selectlanguage
  {english}\bibinfo {title} {Scalar field dark matter as an alternative
  explanation for the anisotropic distribution of satellite galaxies}},\ }\href
  {https://doi.org/10.1103/PhysRevD.103.083535} {\bibfield  {journal} {\bibinfo
   {journal} {Phys. Rev. D}\ }\textbf {\bibinfo {volume} {103}},\ \bibinfo
  {pages} {083535} (\bibinfo {year} {2021})}\BibitemShut {NoStop}%
\bibitem [{\citenamefont {Herdeiro}\ \emph {et~al.}(2021)\citenamefont
  {Herdeiro}, \citenamefont {Pombo}, \citenamefont {Radu}, \citenamefont
  {Cunha},\ and\ \citenamefont {Sanchis-Gual}}]{herdeiro_imitation_2021}%
  \BibitemOpen
  \bibfield  {author} {\bibinfo {author} {\bibfnamefont {C.~A.~R.}\
  \bibnamefont {Herdeiro}}, \bibinfo {author} {\bibfnamefont {A.~M.}\
  \bibnamefont {Pombo}}, \bibinfo {author} {\bibfnamefont {E.}~\bibnamefont
  {Radu}}, \bibinfo {author} {\bibfnamefont {P.~V.~P.}\ \bibnamefont {Cunha}},\
  and\ \bibinfo {author} {\bibfnamefont {N.}~\bibnamefont {Sanchis-Gual}},\
  }\bibfield  {title} {{\selectlanguage {english}\bibinfo {title} {The
  imitation game: {Proca} stars that can mimic the {Schwarzschild} shadow}},\
  }\href {https://doi.org/10.1088/1475-7516/2021/04/051} {\bibfield  {journal}
  {\bibinfo  {journal} {J. Cosmol. Astropart. Phys.}\ }\textbf {\bibinfo
  {volume} {2021}}\bibfield  {number} {\bibinfo  {number} { (04)},\ \bibinfo
  {pages} {051}},\ }\bibinfo {note} {arXiv:2102.01703 [gr-qc,
  physics:hep-th]}\BibitemShut {NoStop}%
\bibitem [{\citenamefont {Olivares}\ \emph {et~al.}(2020)\citenamefont
  {Olivares}, \citenamefont {Younsi}, \citenamefont {Fromm}, \citenamefont
  {De~Laurentis}, \citenamefont {Porth}, \citenamefont {Mizuno}, \citenamefont
  {Falcke}, \citenamefont {Kramer},\ and\ \citenamefont
  {Rezzolla}}]{olivares_how_2020}%
  \BibitemOpen
  \bibfield  {author} {\bibinfo {author} {\bibfnamefont {H.}~\bibnamefont
  {Olivares}}, \bibinfo {author} {\bibfnamefont {Z.}~\bibnamefont {Younsi}},
  \bibinfo {author} {\bibfnamefont {C.~M.}\ \bibnamefont {Fromm}}, \bibinfo
  {author} {\bibfnamefont {M.}~\bibnamefont {De~Laurentis}}, \bibinfo {author}
  {\bibfnamefont {O.}~\bibnamefont {Porth}}, \bibinfo {author} {\bibfnamefont
  {Y.}~\bibnamefont {Mizuno}}, \bibinfo {author} {\bibfnamefont
  {H.}~\bibnamefont {Falcke}}, \bibinfo {author} {\bibfnamefont
  {M.}~\bibnamefont {Kramer}},\ and\ \bibinfo {author} {\bibfnamefont
  {L.}~\bibnamefont {Rezzolla}},\ }\bibfield  {title} {{\selectlanguage
  {english}\bibinfo {title} {How to tell an accreting boson star from a black
  hole}},\ }\href {https://doi.org/10.1093/mnras/staa1878} {\bibfield
  {journal} {\bibinfo  {journal} {Monthly Notices of the Royal Astronomical
  Society}\ }\textbf {\bibinfo {volume} {497}},\ \bibinfo {pages} {521}
  (\bibinfo {year} {2020})},\ \bibinfo {note} {arXiv:1809.08682 [astro-ph,
  physics:gr-qc]}\BibitemShut {NoStop}%
\bibitem [{\citenamefont {Sengo}\ \emph {et~al.}(2023)\citenamefont {Sengo},
  \citenamefont {Cunha}, \citenamefont {Herdeiro},\ and\ \citenamefont
  {Radu}}]{sengo_kerr_2023}%
  \BibitemOpen
  \bibfield  {author} {\bibinfo {author} {\bibfnamefont {I.}~\bibnamefont
  {Sengo}}, \bibinfo {author} {\bibfnamefont {P.~V.~P.}\ \bibnamefont {Cunha}},
  \bibinfo {author} {\bibfnamefont {C.~A.~R.}\ \bibnamefont {Herdeiro}},\ and\
  \bibinfo {author} {\bibfnamefont {E.}~\bibnamefont {Radu}},\ }\bibfield
  {title} {{\selectlanguage {english}\bibinfo {title} {Kerr black holes with
  synchronised {Proca} hair: lensing, shadows and {EHT} constraints}},\ }\href
  {https://doi.org/10.1088/1475-7516/2023/01/047} {\bibfield  {journal}
  {\bibinfo  {journal} {J. Cosmol. Astropart. Phys.}\ }\textbf {\bibinfo
  {volume} {2023}}\bibfield  {number} {\bibinfo  {number} { (01)},\ \bibinfo
  {pages} {047}},\ }\bibinfo {note} {arXiv:2209.06237 [astro-ph,
  physics:gr-qc]}\BibitemShut {NoStop}%
\bibitem [{\citenamefont {Bustillo}\ \emph {et~al.}(2021)\citenamefont
  {Bustillo}, \citenamefont {Sanchis-Gual}, \citenamefont {Torres-Forn{\'e}},
  \citenamefont {Font}, \citenamefont {Vajpeyi}, \citenamefont {Smith},
  \citenamefont {Herdeiro}, \citenamefont {Radu},\ and\ \citenamefont
  {Leong}}]{bustillo_gw190521_2021}%
  \BibitemOpen
  \bibfield  {author} {\bibinfo {author} {\bibfnamefont {J.~C.}\ \bibnamefont
  {Bustillo}}, \bibinfo {author} {\bibfnamefont {N.}~\bibnamefont
  {Sanchis-Gual}}, \bibinfo {author} {\bibfnamefont {A.}~\bibnamefont
  {Torres-Forn{\'e}}}, \bibinfo {author} {\bibfnamefont {J.~A.}\ \bibnamefont
  {Font}}, \bibinfo {author} {\bibfnamefont {A.}~\bibnamefont {Vajpeyi}},
  \bibinfo {author} {\bibfnamefont {R.}~\bibnamefont {Smith}}, \bibinfo
  {author} {\bibfnamefont {C.}~\bibnamefont {Herdeiro}}, \bibinfo {author}
  {\bibfnamefont {E.}~\bibnamefont {Radu}},\ and\ \bibinfo {author}
  {\bibfnamefont {S.~H.~W.}\ \bibnamefont {Leong}},\ }\bibfield  {title}
  {{\selectlanguage {english}\bibinfo {title} {{GW190521} as a merger of
  {Proca} stars: a potential new vector boson of \$8.7 {\textbackslash}times
  10{\textasciicircum}\{-13\}\$ {eV}}},\ }\href
  {https://doi.org/10.1103/PhysRevLett.126.081101} {\bibfield  {journal}
  {\bibinfo  {journal} {Phys. Rev. Lett.}\ }\textbf {\bibinfo {volume} {126}},\
  \bibinfo {pages} {081101} (\bibinfo {year} {2021})},\ \bibinfo {note}
  {arXiv:2009.05376 [astro-ph, physics:gr-qc, physics:hep-ph]}\BibitemShut
  {NoStop}%
\bibitem [{\citenamefont {Visinelli}(2021)}]{visinelli_boson_2021}%
  \BibitemOpen
  \bibfield  {author} {\bibinfo {author} {\bibfnamefont {L.}~\bibnamefont
  {Visinelli}},\ }\bibfield  {title} {{\selectlanguage {english}\bibinfo
  {title} {Boson {Stars} and {Oscillatons}: {A} {Review}}},\ }\href
  {https://doi.org/10.1142/S0218271821300068} {\bibfield  {journal} {\bibinfo
  {journal} {Int. J. Mod. Phys. D}\ }\textbf {\bibinfo {volume} {30}},\
  \bibinfo {pages} {2130006} (\bibinfo {year} {2021})},\ \bibinfo {note}
  {arXiv:2109.05481 [gr-qc, physics:hep-ph]}\BibitemShut {NoStop}%
\bibitem [{\citenamefont {Kaup}(1968)}]{kaup_klein-gordon_1968}%
  \BibitemOpen
  \bibfield  {author} {\bibinfo {author} {\bibfnamefont {D.~J.}\ \bibnamefont
  {Kaup}},\ }\bibfield  {title} {{\selectlanguage {english}\bibinfo {title}
  {Klein-{Gordon} {Geon}}},\ }\href {https://doi.org/10.1103/PhysRev.172.1331}
  {\bibfield  {journal} {\bibinfo  {journal} {Phys. Rev.}\ }\textbf {\bibinfo
  {volume} {172}},\ \bibinfo {pages} {1331} (\bibinfo {year}
  {1968})}\BibitemShut {NoStop}%
\bibitem [{\citenamefont {Brito}\ \emph
  {et~al.}(2016{\natexlab{a}})\citenamefont {Brito}, \citenamefont {Cardoso},
  \citenamefont {Herdeiro},\ and\ \citenamefont {Radu}}]{brito_proca_2016}%
  \BibitemOpen
  \bibfield  {author} {\bibinfo {author} {\bibfnamefont {R.}~\bibnamefont
  {Brito}}, \bibinfo {author} {\bibfnamefont {V.}~\bibnamefont {Cardoso}},
  \bibinfo {author} {\bibfnamefont {C.~A.~R.}\ \bibnamefont {Herdeiro}},\ and\
  \bibinfo {author} {\bibfnamefont {E.}~\bibnamefont {Radu}},\ }\bibfield
  {title} {{\selectlanguage {english}\bibinfo {title} {Proca {Stars}:
  gravitating {Bose}-{Einstein} condensates of massive spin 1 particles}},\
  }\href {https://doi.org/10.1016/j.physletb.2015.11.051} {\bibfield  {journal}
  {\bibinfo  {journal} {Physics Letters B}\ }\textbf {\bibinfo {volume}
  {752}},\ \bibinfo {pages} {291} (\bibinfo {year} {2016}{\natexlab{a}})},\
  \bibinfo {note} {arXiv:1508.05395 [astro-ph, physics:gr-qc, physics:hep-th,
  physics:nlin]}\BibitemShut {NoStop}%
\bibitem [{\citenamefont {Sharma}\ \emph {et~al.}(2008)\citenamefont {Sharma},
  \citenamefont {Karmakar},\ and\ \citenamefont
  {Mukherjee}}]{sharma_boson_2008}%
  \BibitemOpen
  \bibfield  {author} {\bibinfo {author} {\bibfnamefont {R.}~\bibnamefont
  {Sharma}}, \bibinfo {author} {\bibfnamefont {S.}~\bibnamefont {Karmakar}},\
  and\ \bibinfo {author} {\bibfnamefont {S.}~\bibnamefont {Mukherjee}},\ }\href
  {http://arxiv.org/abs/0812.3470} {{\selectlanguage {english}\bibinfo {title}
  {Boson star and dark matter}}} (\bibinfo {year} {2008}),\ \bibinfo {note}
  {arXiv:0812.3470 [gr-qc]}\BibitemShut {NoStop}%
\bibitem [{\citenamefont {Alcubierre}\ \emph {et~al.}(2002)\citenamefont
  {Alcubierre}, \citenamefont {Guzman}, \citenamefont {Matos}, \citenamefont
  {Nunez}, \citenamefont {Urena-Lopez},\ and\ \citenamefont
  {Wiederhold}}]{alcubierre_galactic_2002}%
  \BibitemOpen
  \bibfield  {author} {\bibinfo {author} {\bibfnamefont {M.}~\bibnamefont
  {Alcubierre}}, \bibinfo {author} {\bibfnamefont {F.~S.}\ \bibnamefont
  {Guzman}}, \bibinfo {author} {\bibfnamefont {T.}~\bibnamefont {Matos}},
  \bibinfo {author} {\bibfnamefont {D.}~\bibnamefont {Nunez}}, \bibinfo
  {author} {\bibfnamefont {L.~A.}\ \bibnamefont {Urena-Lopez}},\ and\ \bibinfo
  {author} {\bibfnamefont {P.}~\bibnamefont {Wiederhold}},\ }\bibfield  {title}
  {{\selectlanguage {english}\bibinfo {title} {Galactic {Collapse} of {Scalar}
  {Field} {Dark} {Matter}}},\ }\href
  {https://doi.org/10.1088/0264-9381/19/19/314} {\bibfield  {journal} {\bibinfo
   {journal} {Class. Quantum Grav.}\ }\textbf {\bibinfo {volume} {19}},\
  \bibinfo {pages} {5017} (\bibinfo {year} {2002})},\ \bibinfo {note}
  {arXiv:gr-qc/0110102}\BibitemShut {NoStop}%
\bibitem [{\citenamefont {Herdeiro}(2022)}]{herdeiro_black_2022}%
  \BibitemOpen
  \bibfield  {author} {\bibinfo {author} {\bibfnamefont {C.~A.~R.}\
  \bibnamefont {Herdeiro}},\ }\href {http://arxiv.org/abs/2204.05640}
  {{\selectlanguage {english}\bibinfo {title} {Black holes: on the universality
  of the {Kerr} hypothesis}}} (\bibinfo {year} {2022}),\ \bibinfo {note}
  {arXiv:2204.05640 [astro-ph, physics:gr-qc, physics:hep-th]}\BibitemShut
  {NoStop}%
\bibitem [{\citenamefont {Matos}\ \emph {et~al.}(2022)\citenamefont {Matos},
  \citenamefont {Perez-Lorenzana},\ and\ \citenamefont
  {Sol{\'i}s-L{\'o}pez}}]{matos_fermi_2022}%
  \BibitemOpen
  \bibfield  {author} {\bibinfo {author} {\bibfnamefont {T.}~\bibnamefont
  {Matos}}, \bibinfo {author} {\bibfnamefont {A.}~\bibnamefont
  {Perez-Lorenzana}},\ and\ \bibinfo {author} {\bibfnamefont {J.}~\bibnamefont
  {Sol{\'i}s-L{\'o}pez}},\ }\href {http://arxiv.org/abs/2203.13218}
  {{\selectlanguage {english}\bibinfo {title} {Fermi {Bubbles} in {Scalar}
  {Field} {Dark} {Matter} halos}}} (\bibinfo {year} {2022}),\ \bibinfo {note}
  {arXiv:2203.13218 [astro-ph, physics:gr-qc, physics:hep-ph]}\BibitemShut
  {NoStop}%
\bibitem [{\citenamefont {Kumar}\ \emph {et~al.}(2014)\citenamefont {Kumar},
  \citenamefont {Kulshreshtha},\ and\ \citenamefont
  {Kulshreshtha}}]{kumar_boson_nodate}%
  \BibitemOpen
  \bibfield  {author} {\bibinfo {author} {\bibfnamefont {S.}~\bibnamefont
  {Kumar}}, \bibinfo {author} {\bibfnamefont {U.}~\bibnamefont
  {Kulshreshtha}},\ and\ \bibinfo {author} {\bibfnamefont {D.~S.}\ \bibnamefont
  {Kulshreshtha}},\ }\bibfield  {title} {\bibinfo {title} {Boson stars in a
  theory of complex scalar fields coupled to the u(1) gauge field and
  gravity},\ }\href {https://doi.org/10.1088/0264-9381/31/16/167001} {\bibfield
   {journal} {\bibinfo  {journal} {Classical and Quantum Gravity}\ }\textbf
  {\bibinfo {volume} {31}},\ \bibinfo {pages} {167001} (\bibinfo {year}
  {2014})}\BibitemShut {NoStop}%
\bibitem [{\citenamefont {Matos}\ and\ \citenamefont
  {Castellanos}(2014)}]{matos_phase_2014}%
  \BibitemOpen
  \bibfield  {author} {\bibinfo {author} {\bibfnamefont {T.}~\bibnamefont
  {Matos}}\ and\ \bibinfo {author} {\bibfnamefont {E.}~\bibnamefont
  {Castellanos}},\ }\bibfield  {title} {{\selectlanguage {english}\bibinfo
  {title} {Phase transition from the symmetry breaking of charged
  {Klein}{\textendash}{Gordon} fields}}\ }(\bibinfo {address} {M{\'e}xico City,
  M{\'e}xico},\ \bibinfo {year} {2014})\ pp.\ \bibinfo {pages}
  {181--191}\BibitemShut {NoStop}%
\bibitem [{\citenamefont {Matos}\ \emph {et~al.}(2017)\citenamefont {Matos},
  \citenamefont {Castellanos},\ and\ \citenamefont
  {Su{\'a}rez}}]{matos_bose-einstein_2017}%
  \BibitemOpen
  \bibfield  {author} {\bibinfo {author} {\bibfnamefont {T.}~\bibnamefont
  {Matos}}, \bibinfo {author} {\bibfnamefont {E.}~\bibnamefont {Castellanos}},\
  and\ \bibinfo {author} {\bibfnamefont {A.}~\bibnamefont {Su{\'a}rez}},\
  }\bibfield  {title} {{\selectlanguage {english}\bibinfo {title}
  {Bose-{Einstein} {Condensation} and {Symmetry} {Breaking} of a {Complex}
  {Charged} {Scalar} {Field}}},\ }\href
  {https://doi.org/10.1140/epjc/s10052-017-5065-5} {\bibfield  {journal}
  {\bibinfo  {journal} {Eur. Phys. J. C}\ }\textbf {\bibinfo {volume} {77}},\
  \bibinfo {pages} {500} (\bibinfo {year} {2017})},\ \bibinfo {note}
  {arXiv:1701.04894 [cond-mat]}\BibitemShut {NoStop}%
\bibitem [{\citenamefont {Li}\ \emph {et~al.}(2014)\citenamefont {Li},
  \citenamefont {Rindler-Daller},\ and\ \citenamefont
  {Shapiro}}]{li_cosmological_2014}%
  \BibitemOpen
  \bibfield  {author} {\bibinfo {author} {\bibfnamefont {B.}~\bibnamefont
  {Li}}, \bibinfo {author} {\bibfnamefont {T.}~\bibnamefont {Rindler-Daller}},\
  and\ \bibinfo {author} {\bibfnamefont {P.~R.}\ \bibnamefont {Shapiro}},\
  }\bibfield  {title} {{\selectlanguage {english}\bibinfo {title} {Cosmological
  {Constraints} on {Bose}-{Einstein}-{Condensed} {Scalar} {Field} {Dark}
  {Matter}}},\ }\href {https://doi.org/10.1103/PhysRevD.89.083536} {\bibfield
  {journal} {\bibinfo  {journal} {Phys. Rev. D}\ }\textbf {\bibinfo {volume}
  {89}},\ \bibinfo {pages} {083536} (\bibinfo {year} {2014})},\ \bibinfo {note}
  {arXiv:1310.6061 [astro-ph, physics:gr-qc, physics:hep-ph]}\BibitemShut
  {NoStop}%
\bibitem [{\citenamefont {Herdeiro}\ \emph {et~al.}(2019)\citenamefont
  {Herdeiro}, \citenamefont {Perapechka}, \citenamefont {Radu},\ and\
  \citenamefont {Shnir}}]{herdeiro_asymptotically_2019}%
  \BibitemOpen
  \bibfield  {author} {\bibinfo {author} {\bibfnamefont {C.}~\bibnamefont
  {Herdeiro}}, \bibinfo {author} {\bibfnamefont {I.}~\bibnamefont
  {Perapechka}}, \bibinfo {author} {\bibfnamefont {E.}~\bibnamefont {Radu}},\
  and\ \bibinfo {author} {\bibfnamefont {Y.}~\bibnamefont {Shnir}},\ }\bibfield
   {title} {{\selectlanguage {english}\bibinfo {title} {Asymptotically flat
  spinning scalar, {Dirac} and {Proca} stars}},\ }\href
  {https://doi.org/10.1016/j.physletb.2019.134845} {\bibfield  {journal}
  {\bibinfo  {journal} {Physics Letters B}\ }\textbf {\bibinfo {volume}
  {797}},\ \bibinfo {pages} {134845} (\bibinfo {year} {2019})},\ \bibinfo
  {note} {arXiv:1906.05386 [gr-qc, physics:hep-th]}\BibitemShut {NoStop}%
\bibitem [{\citenamefont {Freitas}\ \emph {et~al.}(2021)\citenamefont
  {Freitas}, \citenamefont {Herdeiro}, \citenamefont {Morais}, \citenamefont
  {Onofre}, \citenamefont {Pasechnik}, \citenamefont {Radu}, \citenamefont
  {Sanchis-Gual},\ and\ \citenamefont {Santos}}]{freitas_ultralight_2021}%
  \BibitemOpen
  \bibfield  {author} {\bibinfo {author} {\bibfnamefont {F.~F.}\ \bibnamefont
  {Freitas}}, \bibinfo {author} {\bibfnamefont {C.~A.~R.}\ \bibnamefont
  {Herdeiro}}, \bibinfo {author} {\bibfnamefont {A.~P.}\ \bibnamefont
  {Morais}}, \bibinfo {author} {\bibfnamefont {A.}~\bibnamefont {Onofre}},
  \bibinfo {author} {\bibfnamefont {R.}~\bibnamefont {Pasechnik}}, \bibinfo
  {author} {\bibfnamefont {E.}~\bibnamefont {Radu}}, \bibinfo {author}
  {\bibfnamefont {N.}~\bibnamefont {Sanchis-Gual}},\ and\ \bibinfo {author}
  {\bibfnamefont {R.}~\bibnamefont {Santos}},\ }\bibfield  {title}
  {{\selectlanguage {english}\bibinfo {title} {Ultralight bosons for strong
  gravity applications from simple {Standard} {Model} extensions}},\ }\href
  {https://doi.org/10.1088/1475-7516/2021/12/047} {\bibfield  {journal}
  {\bibinfo  {journal} {J. Cosmol. Astropart. Phys.}\ }\textbf {\bibinfo
  {volume} {2021}}\bibfield  {number} {\bibinfo  {number} { (12)},\ \bibinfo
  {pages} {047}},\ }\bibinfo {note} {arXiv:2107.09493 [astro-ph, physics:gr-qc,
  physics:hep-ph, physics:hep-th]}\BibitemShut {NoStop}%
\bibitem [{\citenamefont {Caputo}\ \emph {et~al.}(2021)\citenamefont {Caputo},
  \citenamefont {Millar}, \citenamefont {O'Hare},\ and\ \citenamefont
  {Vitagliano}}]{caputo_dark_2021}%
  \BibitemOpen
  \bibfield  {author} {\bibinfo {author} {\bibfnamefont {A.}~\bibnamefont
  {Caputo}}, \bibinfo {author} {\bibfnamefont {A.~J.}\ \bibnamefont {Millar}},
  \bibinfo {author} {\bibfnamefont {C.~A.~J.}\ \bibnamefont {O'Hare}},\ and\
  \bibinfo {author} {\bibfnamefont {E.}~\bibnamefont {Vitagliano}},\ }\bibfield
   {title} {{\selectlanguage {english}\bibinfo {title} {Dark photon limits: a
  handbook}},\ }\href {https://doi.org/10.1103/PhysRevD.104.095029} {\bibfield
  {journal} {\bibinfo  {journal} {Phys. Rev. D}\ }\textbf {\bibinfo {volume}
  {104}},\ \bibinfo {pages} {095029} (\bibinfo {year} {2021})},\ \bibinfo
  {note} {arXiv:2105.04565 [astro-ph, physics:hep-ex, physics:hep-ph,
  physics:physics]}\BibitemShut {NoStop}%
\bibitem [{\citenamefont {Rosa}\ and\ \citenamefont
  {Rubiera-Garcia}(2022)}]{rosa_shadows_2022}%
  \BibitemOpen
  \bibfield  {author} {\bibinfo {author} {\bibfnamefont {J.~L.}\ \bibnamefont
  {Rosa}}\ and\ \bibinfo {author} {\bibfnamefont {D.}~\bibnamefont
  {Rubiera-Garcia}},\ }\bibfield  {title} {{\selectlanguage {english}\bibinfo
  {title} {Shadows of boson and {Proca} stars with thin accretion disks}},\
  }\href {https://doi.org/10.1103/PhysRevD.106.084004} {\bibfield  {journal}
  {\bibinfo  {journal} {Phys. Rev. D}\ }\textbf {\bibinfo {volume} {106}},\
  \bibinfo {pages} {084004} (\bibinfo {year} {2022})},\ \bibinfo {note}
  {arXiv:2204.12949 [astro-ph, physics:gr-qc]}\BibitemShut {NoStop}%
\bibitem [{\citenamefont {Bolton}\ \emph {et~al.}(2022)\citenamefont {Bolton},
  \citenamefont {Caputo}, \citenamefont {Liu},\ and\ \citenamefont
  {Viel}}]{bolton_comparison_2022}%
  \BibitemOpen
  \bibfield  {author} {\bibinfo {author} {\bibfnamefont {J.~S.}\ \bibnamefont
  {Bolton}}, \bibinfo {author} {\bibfnamefont {A.}~\bibnamefont {Caputo}},
  \bibinfo {author} {\bibfnamefont {H.}~\bibnamefont {Liu}},\ and\ \bibinfo
  {author} {\bibfnamefont {M.}~\bibnamefont {Viel}},\ }\bibfield  {title}
  {{\selectlanguage {english}\bibinfo {title} {Comparison of {Low}-{Redshift}
  {Lyman}- $\alpha$ {Forest} {Observations} to {Hydrodynamical} {Simulations}
  with {Dark} {Photon} {Dark} {Matter}}},\ }\href
  {https://doi.org/10.1103/PhysRevLett.129.211102} {\bibfield  {journal}
  {\bibinfo  {journal} {Phys. Rev. Lett.}\ }\textbf {\bibinfo {volume} {129}},\
  \bibinfo {pages} {211102} (\bibinfo {year} {2022})}\BibitemShut {NoStop}%
\bibitem [{\citenamefont {Ji}\ and\ \citenamefont
  {Sin}(1994{\natexlab{b}})}]{ji_late-time_1994-1}%
  \BibitemOpen
  \bibfield  {author} {\bibinfo {author} {\bibfnamefont {S.~U.}\ \bibnamefont
  {Ji}}\ and\ \bibinfo {author} {\bibfnamefont {S.~J.}\ \bibnamefont {Sin}},\
  }\bibfield  {title} {{\selectlanguage {english}\bibinfo {title} {Late-time
  phase transition and the galactic halo as a {Bose} liquid. {II}. {The} effect
  of visible matter}},\ }\href {https://doi.org/10.1103/PhysRevD.50.3655}
  {\bibfield  {journal} {\bibinfo  {journal} {Phys. Rev. D}\ }\textbf {\bibinfo
  {volume} {50}},\ \bibinfo {pages} {3655} (\bibinfo {year}
  {1994}{\natexlab{b}})}\BibitemShut {NoStop}%
\bibitem [{\citenamefont {Matos}\ \emph {et~al.}(2000)\citenamefont {Matos},
  \citenamefont {Guzman},\ and\ \citenamefont
  {Urena-Lopez}}]{matos_scalar_2000}%
  \BibitemOpen
  \bibfield  {author} {\bibinfo {author} {\bibfnamefont {T.}~\bibnamefont
  {Matos}}, \bibinfo {author} {\bibfnamefont {F.~S.}\ \bibnamefont {Guzman}},\
  and\ \bibinfo {author} {\bibfnamefont {L.~A.}\ \bibnamefont {Urena-Lopez}},\
  }\bibfield  {title} {{\selectlanguage {english}\bibinfo {title} {Scalar
  {Field} as {Dark} {Matter} in the {Universe}}},\ }\href
  {https://doi.org/10.1088/0264-9381/17/7/309} {\bibfield  {journal} {\bibinfo
  {journal} {Class. Quantum Grav.}\ }\textbf {\bibinfo {volume} {17}},\
  \bibinfo {pages} {1707} (\bibinfo {year} {2000})},\ \bibinfo {note}
  {arXiv:astro-ph/9908152}\BibitemShut {NoStop}%
\bibitem [{\citenamefont {Matos}\ and\ \citenamefont
  {Urena-Lopez}(2000)}]{matos_quintessence_2000}%
  \BibitemOpen
  \bibfield  {author} {\bibinfo {author} {\bibfnamefont {T.}~\bibnamefont
  {Matos}}\ and\ \bibinfo {author} {\bibfnamefont {L.~A.}\ \bibnamefont
  {Urena-Lopez}},\ }\bibfield  {title} {{\selectlanguage {english}\bibinfo
  {title} {Quintessence and {Scalar} {Dark} {Matter} in the {Universe}}},\
  }\href {https://doi.org/10.1088/0264-9381/17/13/101} {\bibfield  {journal}
  {\bibinfo  {journal} {Class. Quantum Grav.}\ }\textbf {\bibinfo {volume}
  {17}},\ \bibinfo {pages} {L75} (\bibinfo {year} {2000})},\ \bibinfo {note}
  {arXiv:astro-ph/0004332}\BibitemShut {NoStop}%
\bibitem [{\citenamefont {Sahni}\ and\ \citenamefont
  {Wang}(2000)}]{sahni_new_2000}%
  \BibitemOpen
  \bibfield  {author} {\bibinfo {author} {\bibfnamefont {V.}~\bibnamefont
  {Sahni}}\ and\ \bibinfo {author} {\bibfnamefont {L.}~\bibnamefont {Wang}},\
  }\bibfield  {title} {{\selectlanguage {english}\bibinfo {title} {A {New}
  {Cosmological} {Model} of {Quintessence} and {Dark} {Matter}}},\ }\href
  {https://doi.org/10.1103/PhysRevD.62.103517} {\bibfield  {journal} {\bibinfo
  {journal} {Phys. Rev. D}\ }\textbf {\bibinfo {volume} {62}},\ \bibinfo
  {pages} {103517} (\bibinfo {year} {2000})},\ \bibinfo {note}
  {arXiv:astro-ph/9910097}\BibitemShut {NoStop}%
\bibitem [{\citenamefont {Matos}\ \emph {et~al.}(2010)\citenamefont {Matos},
  \citenamefont {Su{\'a}rez}, \citenamefont {Maga{\~n}a}, \citenamefont
  {{Instituto Avanzado de Cosmolog{\'i}a (IAC) collaboration}}, \citenamefont
  {Morales-Tecotl}, \citenamefont {Urena-Lopez}, \citenamefont
  {Linares-Romero},\ and\ \citenamefont
  {Garcia-Compean}}]{matos_structure_2010}%
  \BibitemOpen
  \bibfield  {author} {\bibinfo {author} {\bibfnamefont {T.}~\bibnamefont
  {Matos}}, \bibinfo {author} {\bibfnamefont {A.}~\bibnamefont {Su{\'a}rez}},
  \bibinfo {author} {\bibfnamefont {J.~A.}\ \bibnamefont {Maga{\~n}a}},
  \bibinfo {author} {\bibnamefont {{Instituto Avanzado de Cosmolog{\'i}a (IAC)
  collaboration}}}, \bibinfo {author} {\bibfnamefont {H.~A.}\ \bibnamefont
  {Morales-Tecotl}}, \bibinfo {author} {\bibfnamefont {L.~A.}\ \bibnamefont
  {Urena-Lopez}}, \bibinfo {author} {\bibfnamefont {R.}~\bibnamefont
  {Linares-Romero}},\ and\ \bibinfo {author} {\bibfnamefont {H.~H.}\
  \bibnamefont {Garcia-Compean}},\ }\bibfield  {title} {{\selectlanguage
  {english}\bibinfo {title} {Structure {Formation} with {Scalar} {Field} {Dark}
  {Matter}}}\ }(\bibinfo {year} {2010})\ pp.\ \bibinfo {pages}
  {283--287}\BibitemShut {NoStop}%
\bibitem [{\citenamefont {Turner}(1983)}]{turner_coherent_1983}%
  \BibitemOpen
  \bibfield  {author} {\bibinfo {author} {\bibfnamefont {M.~S.}\ \bibnamefont
  {Turner}},\ }\bibfield  {title} {{\selectlanguage {english}\bibinfo {title}
  {Coherent scalar-field oscillations in an expanding universe}},\ }\href
  {https://doi.org/10.1103/PhysRevD.28.1243} {\bibfield  {journal} {\bibinfo
  {journal} {Phys. Rev. D}\ }\textbf {\bibinfo {volume} {28}},\ \bibinfo
  {pages} {1243} (\bibinfo {year} {1983})}\BibitemShut {NoStop}%
\bibitem [{\citenamefont {Briscese}(2011)}]{briscese_viability_2011}%
  \BibitemOpen
  \bibfield  {author} {\bibinfo {author} {\bibfnamefont {F.}~\bibnamefont
  {Briscese}},\ }\bibfield  {title} {{\selectlanguage {english}\bibinfo {title}
  {Viability of complex self-interacting scalar field as dark matter}},\ }\href
  {https://doi.org/10.1016/j.physletb.2010.12.064} {\bibfield  {journal}
  {\bibinfo  {journal} {Physics Letters B}\ }\textbf {\bibinfo {volume}
  {696}},\ \bibinfo {pages} {315} (\bibinfo {year} {2011})},\ \bibinfo {note}
  {arXiv:1101.0028 [astro-ph, physics:gr-qc, physics:hep-ph]}\BibitemShut
  {NoStop}%
\bibitem [{\citenamefont {Su}\ \emph {et~al.}(2010)\citenamefont {Su},
  \citenamefont {Slatyer},\ and\ \citenamefont {Finkbeiner}}]{su_giant_2010}%
  \BibitemOpen
  \bibfield  {author} {\bibinfo {author} {\bibfnamefont {M.}~\bibnamefont
  {Su}}, \bibinfo {author} {\bibfnamefont {T.~R.}\ \bibnamefont {Slatyer}},\
  and\ \bibinfo {author} {\bibfnamefont {D.~P.}\ \bibnamefont {Finkbeiner}},\
  }\bibfield  {title} {{\selectlanguage {english}\bibinfo {title} {Giant
  {Gamma}-ray {Bubbles} from {Fermi}-{LAT}: {AGN} {Activity} or {Bipolar}
  {Galactic} {Wind}?}},\ }\href {https://doi.org/10.1088/0004-637X/724/2/1044}
  {\bibfield  {journal} {\bibinfo  {journal} {ApJ}\ }\textbf {\bibinfo {volume}
  {724}},\ \bibinfo {pages} {1044} (\bibinfo {year} {2010})},\ \bibinfo {note}
  {arXiv:1005.5480 [astro-ph]}\BibitemShut {NoStop}%
\bibitem [{\citenamefont {Collaboration}(2014)}]{collaboration_spectrum_2014}%
  \BibitemOpen
  \bibfield  {author} {\bibinfo {author} {\bibfnamefont {F.-L.}\ \bibnamefont
  {Collaboration}},\ }\bibfield  {title} {{\selectlanguage {english}\bibinfo
  {title} {The {Spectrum} and {Morphology} of the {Fermi} {Bubbles}}},\ }\href
  {https://doi.org/10.1088/0004-637X/793/1/64} {\bibfield  {journal} {\bibinfo
  {journal} {ApJ}\ }\textbf {\bibinfo {volume} {793}},\ \bibinfo {pages} {64}
  (\bibinfo {year} {2014})},\ \bibinfo {note} {arXiv:1407.7905
  [astro-ph]}\BibitemShut {NoStop}%
\bibitem [{\citenamefont {Arbey}\ and\ \citenamefont
  {Mahmoudi}(2021)}]{arbey_dark_2021}%
  \BibitemOpen
  \bibfield  {author} {\bibinfo {author} {\bibfnamefont {A.}~\bibnamefont
  {Arbey}}\ and\ \bibinfo {author} {\bibfnamefont {F.}~\bibnamefont
  {Mahmoudi}},\ }\bibfield  {title} {{\selectlanguage {english}\bibinfo {title}
  {Dark matter and the early {Universe}: a review}},\ }\href
  {https://doi.org/10.1016/j.ppnp.2021.103865} {\bibfield  {journal} {\bibinfo
  {journal} {Progress in Particle and Nuclear Physics}\ }\textbf {\bibinfo
  {volume} {119}},\ \bibinfo {pages} {103865} (\bibinfo {year} {2021})},\
  \bibinfo {note} {arXiv:2104.11488 [astro-ph, physics:hep-ph]}\BibitemShut
  {NoStop}%
\bibitem [{\citenamefont {Freese}(2009)}]{freese_review_2009}%
  \BibitemOpen
  \bibfield  {author} {\bibinfo {author} {\bibfnamefont {K.}~\bibnamefont
  {Freese}},\ }\bibfield  {title} {{\selectlanguage {english}\bibinfo {title}
  {Review of {Observational} {Evidence} for {Dark} {Matter} in the {Universe}
  and in upcoming searches for {Dark} {Stars}}},\ }\href
  {https://doi.org/10.1051/eas/0936016} {\bibfield  {journal} {\bibinfo
  {journal} {EAS Publications Series}\ }\textbf {\bibinfo {volume} {36}},\
  \bibinfo {pages} {113} (\bibinfo {year} {2009})},\ \bibinfo {note}
  {arXiv:0812.4005 [astro-ph, physics:hep-ph]}\BibitemShut {NoStop}%
\bibitem [{\citenamefont {Khlopov}(2017)}]{khlopov_particle_2017}%
  \BibitemOpen
  \bibfield  {author} {\bibinfo {author} {\bibfnamefont {M.~Y.}\ \bibnamefont
  {Khlopov}},\ }\bibfield  {title} {{\selectlanguage {english}\bibinfo {title}
  {Particle {Dark} {Matter} {Candidates}}},\ }\href
  {https://doi.org/10.1142/S0217732317020011} {\bibfield  {journal} {\bibinfo
  {journal} {Mod. Phys. Lett. A}\ }\textbf {\bibinfo {volume} {32}},\ \bibinfo
  {pages} {1702001} (\bibinfo {year} {2017})},\ \bibinfo {note}
  {arXiv:1704.06511 [hep-ph]}\BibitemShut {NoStop}%
\bibitem [{\citenamefont {Ryder}(1996)}]{ryder_quantum_1996}%
  \BibitemOpen
  \bibfield  {author} {\bibinfo {author} {\bibfnamefont {L.~H.}\ \bibnamefont
  {Ryder}},\ }\href {https://doi.org/10.1017/CBO9780511813900}
  {{\selectlanguage {english}\emph {\bibinfo {title} {Quantum {Field}
  {Theory}}}}},\ \bibinfo {edition} {2nd}\ ed.\ (\bibinfo  {publisher}
  {Cambridge University Press},\ \bibinfo {year} {1996})\BibitemShut {NoStop}%
\bibitem [{\citenamefont {Gan}(2019)}]{gan_gauge_2019}%
  \BibitemOpen
  \bibfield  {author} {\bibinfo {author} {\bibfnamefont {W.~S.}\ \bibnamefont
  {Gan}},\ }\href {https://doi.org/10.1007/978-981-13-8751-7} {{\selectlanguage
  {english}\emph {\bibinfo {title} {Gauge {Invariance} {Approach} to {Acoustic}
  {Fields}}}}}\ (\bibinfo  {publisher} {Springer Singapore},\ \bibinfo
  {address} {Singapore},\ \bibinfo {year} {2019})\BibitemShut {NoStop}%
\bibitem [{\citenamefont {Lancaster}\ and\ \citenamefont
  {Blundell}(2014)}]{lancaster_quantum_2014}%
  \BibitemOpen
  \bibfield  {author} {\bibinfo {author} {\bibfnamefont {T.}~\bibnamefont
  {Lancaster}}\ and\ \bibinfo {author} {\bibfnamefont {S.}~\bibnamefont
  {Blundell}},\ }\href@noop {} {{\selectlanguage {english}\emph {\bibinfo
  {title} {Quantum field theory for the gifted amateur}}}},\ \bibinfo {edition}
  {first edition}\ ed.\ (\bibinfo  {publisher} {Oxford University Press},\
  \bibinfo {address} {Oxford},\ \bibinfo {year} {2014})\ \bibinfo {note} {oCLC:
  ocn859651399}\BibitemShut {NoStop}%
\bibitem [{\citenamefont {Schunck}\ and\ \citenamefont
  {Mielke}(2003)}]{schunck_topical_2003}%
  \BibitemOpen
  \bibfield  {author} {\bibinfo {author} {\bibfnamefont {F.~E.}\ \bibnamefont
  {Schunck}}\ and\ \bibinfo {author} {\bibfnamefont {E.~W.}\ \bibnamefont
  {Mielke}},\ }\bibfield  {title} {{\selectlanguage {english}\bibinfo {title}
  {{TOPICAL} {REVIEW}: {General} relativistic boson stars}},\ }\href
  {https://doi.org/10.1088/0264-9381/20/20/201} {\bibfield  {journal} {\bibinfo
   {journal} {Class. Quantum Grav.}\ }\textbf {\bibinfo {volume} {20}},\
  \bibinfo {pages} {R301} (\bibinfo {year} {2003})},\ \bibinfo {note}
  {arXiv:0801.0307 [astro-ph]}\BibitemShut {NoStop}%
\bibitem [{\citenamefont {Ruffini}\ and\ \citenamefont
  {Bonazzola}(1969)}]{ruffini_systems_1969}%
  \BibitemOpen
  \bibfield  {author} {\bibinfo {author} {\bibfnamefont {R.}~\bibnamefont
  {Ruffini}}\ and\ \bibinfo {author} {\bibfnamefont {S.}~\bibnamefont
  {Bonazzola}},\ }\bibfield  {title} {{\selectlanguage {english}\bibinfo
  {title} {Systems of {Self}-{Gravitating} {Particles} in {General}
  {Relativity} and the {Concept} of an {Equation} of {State}}},\ }\href
  {https://doi.org/10.1103/PhysRev.187.1767} {\bibfield  {journal} {\bibinfo
  {journal} {Phys. Rev.}\ }\textbf {\bibinfo {volume} {187}},\ \bibinfo {pages}
  {1767} (\bibinfo {year} {1969})}\BibitemShut {NoStop}%
\bibitem [{\citenamefont {Collaboration}(2022)}]{antypas_new_2022}%
  \BibitemOpen
  \bibfield  {author} {\bibinfo {author} {\bibnamefont {Collaboration}},\
  }\href {http://arxiv.org/abs/2203.14915} {{\selectlanguage {english}\bibinfo
  {title} {New {Horizons}: {Scalar} and {Vector} {Ultralight} {Dark}
  {Matter}}}} (\bibinfo {year} {2022}),\ \bibinfo {note} {arXiv:2203.14915
  [astro-ph, physics:hep-ex, physics:hep-ph, physics:physics,
  physics:quant-ph]}\BibitemShut {NoStop}%
\bibitem [{\citenamefont {Meliani}\ \emph {et~al.}(2015)\citenamefont
  {Meliani}, \citenamefont {Vincent}, \citenamefont {Grandcl{\'e}ment},
  \citenamefont {Gourgoulhon}, \citenamefont {Monceau-Baroux},\ and\
  \citenamefont {Straub}}]{meliani_circular_2015}%
  \BibitemOpen
  \bibfield  {author} {\bibinfo {author} {\bibfnamefont {Z.}~\bibnamefont
  {Meliani}}, \bibinfo {author} {\bibfnamefont {F.~H.}\ \bibnamefont
  {Vincent}}, \bibinfo {author} {\bibfnamefont {P.}~\bibnamefont
  {Grandcl{\'e}ment}}, \bibinfo {author} {\bibfnamefont {E.}~\bibnamefont
  {Gourgoulhon}}, \bibinfo {author} {\bibfnamefont {R.}~\bibnamefont
  {Monceau-Baroux}},\ and\ \bibinfo {author} {\bibfnamefont {O.}~\bibnamefont
  {Straub}},\ }\bibfield  {title} {{\selectlanguage {english}\bibinfo {title}
  {Circular geodesics and thick tori around rotating boson stars}},\ }\href
  {https://doi.org/10.1088/0264-9381/32/23/235022} {\bibfield  {journal}
  {\bibinfo  {journal} {Class. Quantum Grav.}\ }\textbf {\bibinfo {volume}
  {32}},\ \bibinfo {pages} {235022} (\bibinfo {year} {2015})},\ \bibinfo {note}
  {arXiv:1510.04191 [astro-ph, physics:gr-qc]}\BibitemShut {NoStop}%
\bibitem [{\citenamefont {Grandclement}\ \emph {et~al.}(2014)\citenamefont
  {Grandclement}, \citenamefont {Som{\'e}},\ and\ \citenamefont
  {Gourgoulhon}}]{grandclement_models_2014}%
  \BibitemOpen
  \bibfield  {author} {\bibinfo {author} {\bibfnamefont {P.}~\bibnamefont
  {Grandclement}}, \bibinfo {author} {\bibfnamefont {C.}~\bibnamefont
  {Som{\'e}}},\ and\ \bibinfo {author} {\bibfnamefont {E.}~\bibnamefont
  {Gourgoulhon}},\ }\bibfield  {title} {{\selectlanguage {english}\bibinfo
  {title} {Models of rotating boson stars and geodesics around them: new type
  of orbits}},\ }\href {https://doi.org/10.1103/PhysRevD.90.024068} {\bibfield
  {journal} {\bibinfo  {journal} {Phys. Rev. D}\ }\textbf {\bibinfo {volume}
  {90}},\ \bibinfo {pages} {024068} (\bibinfo {year} {2014})},\ \bibinfo {note}
  {arXiv:1405.4837 [gr-qc]}\BibitemShut {NoStop}%
\bibitem [{\citenamefont {Vincent}\ \emph {et~al.}(2016)\citenamefont
  {Vincent}, \citenamefont {Meliani}, \citenamefont {Grandclement},
  \citenamefont {Gourgoulhon},\ and\ \citenamefont
  {Straub}}]{vincent_imaging_2016}%
  \BibitemOpen
  \bibfield  {author} {\bibinfo {author} {\bibfnamefont {F.~H.}\ \bibnamefont
  {Vincent}}, \bibinfo {author} {\bibfnamefont {Z.}~\bibnamefont {Meliani}},
  \bibinfo {author} {\bibfnamefont {P.}~\bibnamefont {Grandclement}}, \bibinfo
  {author} {\bibfnamefont {E.}~\bibnamefont {Gourgoulhon}},\ and\ \bibinfo
  {author} {\bibfnamefont {O.}~\bibnamefont {Straub}},\ }\bibfield  {title}
  {{\selectlanguage {english}\bibinfo {title} {Imaging a boson star at the
  {Galactic} center}},\ }\href {https://doi.org/10.1088/0264-9381/33/10/105015}
  {\bibfield  {journal} {\bibinfo  {journal} {Class. Quantum Grav.}\ }\textbf
  {\bibinfo {volume} {33}},\ \bibinfo {pages} {105015} (\bibinfo {year}
  {2016})},\ \bibinfo {note} {arXiv:1510.04170 [astro-ph,
  physics:gr-qc]}\BibitemShut {NoStop}%
\bibitem [{\citenamefont {Cunha}\ \emph {et~al.}(2016)\citenamefont {Cunha},
  \citenamefont {Herdeiro}, \citenamefont {Radu},\ and\ \citenamefont
  {Runarsson}}]{cunha_shadows_2016}%
  \BibitemOpen
  \bibfield  {author} {\bibinfo {author} {\bibfnamefont {P.~V.~P.}\
  \bibnamefont {Cunha}}, \bibinfo {author} {\bibfnamefont {C.~A.~R.}\
  \bibnamefont {Herdeiro}}, \bibinfo {author} {\bibfnamefont {E.}~\bibnamefont
  {Radu}},\ and\ \bibinfo {author} {\bibfnamefont {H.~F.}\ \bibnamefont
  {Runarsson}},\ }\bibfield  {title} {{\selectlanguage {english}\bibinfo
  {title} {Shadows of {Kerr} black holes with and without scalar hair}},\
  }\href {https://doi.org/10.1142/S0218271816410212} {\bibfield  {journal}
  {\bibinfo  {journal} {Int. J. Mod. Phys. D}\ }\textbf {\bibinfo {volume}
  {25}},\ \bibinfo {pages} {1641021} (\bibinfo {year} {2016})},\ \bibinfo
  {note} {arXiv:1605.08293 [astro-ph, physics:gr-qc,
  physics:hep-th]}\BibitemShut {NoStop}%
\bibitem [{\citenamefont {Dehnen}\ and\ \citenamefont
  {Rose}(1993)}]{dehnen_flat_1993}%
  \BibitemOpen
  \bibfield  {author} {\bibinfo {author} {\bibfnamefont {H.}~\bibnamefont
  {Dehnen}}\ and\ \bibinfo {author} {\bibfnamefont {B.}~\bibnamefont {Rose}},\
  }\bibfield  {title} {{\selectlanguage {english}\bibinfo {title} {Flat
  rotation curves of spiral galaxies and the dark matter particles}},\ }\href
  {https://doi.org/10.1007/BF00659137} {\bibfield  {journal} {\bibinfo
  {journal} {Astrophys Space Sci}\ }\textbf {\bibinfo {volume} {207}},\
  \bibinfo {pages} {133} (\bibinfo {year} {1993})}\BibitemShut {NoStop}%
\bibitem [{\citenamefont {Garc{\'i}a}\ and\ \citenamefont
  {Landea}(2016)}]{garcia_charged_2016}%
  \BibitemOpen
  \bibfield  {author} {\bibinfo {author} {\bibfnamefont {F.}~\bibnamefont
  {Garc{\'i}a}}\ and\ \bibinfo {author} {\bibfnamefont {I.~S.}\ \bibnamefont
  {Landea}},\ }\bibfield  {title} {{\selectlanguage {english}\bibinfo {title}
  {Charged {Proca} {Stars}}},\ }\href
  {https://doi.org/10.1103/PhysRevD.94.104006} {\bibfield  {journal} {\bibinfo
  {journal} {Phys. Rev. D}\ }\textbf {\bibinfo {volume} {94}},\ \bibinfo
  {pages} {104006} (\bibinfo {year} {2016})},\ \bibinfo {note}
  {arXiv:1608.00011 [astro-ph, physics:gr-qc, physics:hep-th]}\BibitemShut
  {NoStop}%
\bibitem [{\citenamefont {Minamitsuji}(2018)}]{minamitsuji_vector_2018}%
  \BibitemOpen
  \bibfield  {author} {\bibinfo {author} {\bibfnamefont {M.}~\bibnamefont
  {Minamitsuji}},\ }\bibfield  {title} {{\selectlanguage {english}\bibinfo
  {title} {Vector boson star solutions with a quartic order
  self-interaction}},\ }\href {https://doi.org/10.1103/PhysRevD.97.104023}
  {\bibfield  {journal} {\bibinfo  {journal} {Phys. Rev. D}\ }\textbf {\bibinfo
  {volume} {97}},\ \bibinfo {pages} {104023} (\bibinfo {year}
  {2018})}\BibitemShut {NoStop}%
\bibitem [{\citenamefont {Brito}\ \emph
  {et~al.}(2016{\natexlab{b}})\citenamefont {Brito}, \citenamefont {Cardoso},
  \citenamefont {Macedo}, \citenamefont {Okawa},\ and\ \citenamefont
  {Palenzuela}}]{brito_interaction_2016}%
  \BibitemOpen
  \bibfield  {author} {\bibinfo {author} {\bibfnamefont {R.}~\bibnamefont
  {Brito}}, \bibinfo {author} {\bibfnamefont {V.}~\bibnamefont {Cardoso}},
  \bibinfo {author} {\bibfnamefont {C.~F.~B.}\ \bibnamefont {Macedo}}, \bibinfo
  {author} {\bibfnamefont {H.}~\bibnamefont {Okawa}},\ and\ \bibinfo {author}
  {\bibfnamefont {C.}~\bibnamefont {Palenzuela}},\ }\bibfield  {title}
  {{\selectlanguage {english}\bibinfo {title} {Interaction between bosonic dark
  matter and stars}},\ }\href {https://doi.org/10.1103/PhysRevD.93.044045}
  {\bibfield  {journal} {\bibinfo  {journal} {Phys. Rev. D}\ }\textbf {\bibinfo
  {volume} {93}},\ \bibinfo {pages} {044045} (\bibinfo {year}
  {2016}{\natexlab{b}})},\ \bibinfo {note} {arXiv:1512.00466 [astro-ph,
  physics:gr-qc, physics:hep-ph]}\BibitemShut {NoStop}%
\bibitem [{\citenamefont {Jaeckel}(2013)}]{jaeckel_force_2013}%
  \BibitemOpen
  \bibfield  {author} {\bibinfo {author} {\bibfnamefont {J.}~\bibnamefont
  {Jaeckel}},\ }\href {http://arxiv.org/abs/1303.1821} {{\selectlanguage
  {english}\bibinfo {title} {A force beyond the {Standard} {Model} - {Status}
  of the quest for hidden photons}}} (\bibinfo {year} {2013}),\ \bibinfo {note}
  {arXiv:1303.1821 [hep-ph]}\BibitemShut {NoStop}%
\bibitem [{\citenamefont {Sato}\ \emph {et~al.}(2022)\citenamefont {Sato},
  \citenamefont {Takahashi},\ and\ \citenamefont
  {Yamada}}]{sato_gravitational_2022}%
  \BibitemOpen
  \bibfield  {author} {\bibinfo {author} {\bibfnamefont {T.}~\bibnamefont
  {Sato}}, \bibinfo {author} {\bibfnamefont {F.}~\bibnamefont {Takahashi}},\
  and\ \bibinfo {author} {\bibfnamefont {M.}~\bibnamefont {Yamada}},\
  }\bibfield  {title} {{\selectlanguage {english}\bibinfo {title}
  {Gravitational production of dark photon dark matter with mass generated by
  the {Higgs} mechanism}},\ }\href
  {https://doi.org/10.1088/1475-7516/2022/08/022} {\bibfield  {journal}
  {\bibinfo  {journal} {J. Cosmol. Astropart. Phys.}\ }\textbf {\bibinfo
  {volume} {2022}}\bibfield  {number} {\bibinfo  {number} { (08)},\ \bibinfo
  {pages} {022}},\ }\bibinfo {note} {arXiv:2204.11896 [astro-ph,
  physics:hep-ph]}\BibitemShut {NoStop}%
\bibitem [{\citenamefont {Alonso-{\'A}lvarez}\ \emph
  {et~al.}(2020)\citenamefont {Alonso-{\'A}lvarez}, \citenamefont {Ertas},
  \citenamefont {Jaeckel}, \citenamefont {Kahlhoefer},\ and\ \citenamefont
  {Thormaehlen}}]{alonso-alvarez_hidden_2020}%
  \BibitemOpen
  \bibfield  {author} {\bibinfo {author} {\bibfnamefont {G.}~\bibnamefont
  {Alonso-{\'A}lvarez}}, \bibinfo {author} {\bibfnamefont {F.}~\bibnamefont
  {Ertas}}, \bibinfo {author} {\bibfnamefont {J.}~\bibnamefont {Jaeckel}},
  \bibinfo {author} {\bibfnamefont {F.}~\bibnamefont {Kahlhoefer}},\ and\
  \bibinfo {author} {\bibfnamefont {L.~J.}\ \bibnamefont {Thormaehlen}},\
  }\bibfield  {title} {{\selectlanguage {english}\bibinfo {title} {Hidden
  {Photon} {Dark} {Matter} in the {Light} of {XENON1T} and {Stellar}
  {Cooling}}},\ }\href {https://doi.org/10.1088/1475-7516/2020/11/029}
  {\bibfield  {journal} {\bibinfo  {journal} {J. Cosmol. Astropart. Phys.}\
  }\textbf {\bibinfo {volume} {2020}}\bibfield  {number} {\bibinfo  {number} {
  (11)},\ \bibinfo {pages} {029}},\ }\bibinfo {note} {arXiv:2006.11243
  [astro-ph, physics:hep-ex, physics:hep-ph]}\BibitemShut {NoStop}%
\bibitem [{\citenamefont {Agrawal}\ \emph {et~al.}(2020)\citenamefont
  {Agrawal}, \citenamefont {Kitajima}, \citenamefont {Reece}, \citenamefont
  {Sekiguchi},\ and\ \citenamefont {Takahashi}}]{agrawal_relic_2020}%
  \BibitemOpen
  \bibfield  {author} {\bibinfo {author} {\bibfnamefont {P.}~\bibnamefont
  {Agrawal}}, \bibinfo {author} {\bibfnamefont {N.}~\bibnamefont {Kitajima}},
  \bibinfo {author} {\bibfnamefont {M.}~\bibnamefont {Reece}}, \bibinfo
  {author} {\bibfnamefont {T.}~\bibnamefont {Sekiguchi}},\ and\ \bibinfo
  {author} {\bibfnamefont {F.}~\bibnamefont {Takahashi}},\ }\bibfield  {title}
  {{\selectlanguage {english}\bibinfo {title} {Relic {Abundance} of {Dark}
  {Photon} {Dark} {Matter}}},\ }\href
  {https://doi.org/10.1016/j.physletb.2019.135136} {\bibfield  {journal}
  {\bibinfo  {journal} {Physics Letters B}\ }\textbf {\bibinfo {volume}
  {801}},\ \bibinfo {pages} {135136} (\bibinfo {year} {2020})},\ \bibinfo
  {note} {arXiv:1810.07188 [hep-ph]}\BibitemShut {NoStop}%
\bibitem [{\citenamefont {Adshead}\ and\ \citenamefont
  {Lozanov}(2021)}]{adshead_self-gravitating_2021}%
  \BibitemOpen
  \bibfield  {author} {\bibinfo {author} {\bibfnamefont {P.}~\bibnamefont
  {Adshead}}\ and\ \bibinfo {author} {\bibfnamefont {K.~D.}\ \bibnamefont
  {Lozanov}},\ }\bibfield  {title} {{\selectlanguage {english}\bibinfo {title}
  {Self-gravitating {Vector} {Dark} {Matter}}},\ }\href
  {https://doi.org/10.1103/PhysRevD.103.103501} {\bibfield  {journal} {\bibinfo
   {journal} {Phys. Rev. D}\ }\textbf {\bibinfo {volume} {103}},\ \bibinfo
  {pages} {103501} (\bibinfo {year} {2021})},\ \bibinfo {note}
  {arXiv:2101.07265 [astro-ph, physics:gr-qc, physics:hep-ph]}\BibitemShut
  {NoStop}%
\bibitem [{\citenamefont {Amin}\ \emph {et~al.}(2022)\citenamefont {Amin},
  \citenamefont {Jain}, \citenamefont {Karur},\ and\ \citenamefont
  {Mocz}}]{amin_small-scale_2022}%
  \BibitemOpen
  \bibfield  {author} {\bibinfo {author} {\bibfnamefont {M.~A.}\ \bibnamefont
  {Amin}}, \bibinfo {author} {\bibfnamefont {M.}~\bibnamefont {Jain}}, \bibinfo
  {author} {\bibfnamefont {R.}~\bibnamefont {Karur}},\ and\ \bibinfo {author}
  {\bibfnamefont {P.}~\bibnamefont {Mocz}},\ }\bibfield  {title}
  {{\selectlanguage {english}\bibinfo {title} {Small-scale structure in vector
  dark matter}},\ }\href {https://doi.org/10.1088/1475-7516/2022/08/014}
  {\bibfield  {journal} {\bibinfo  {journal} {J. Cosmol. Astropart. Phys.}\
  }\textbf {\bibinfo {volume} {2022}}\bibfield  {number} {\bibinfo  {number} {
  (08)},\ \bibinfo {pages} {014}},\ }\bibinfo {note} {arXiv:2203.11935
  [astro-ph, physics:hep-ph]}\BibitemShut {NoStop}%
\bibitem [{\citenamefont {Denig}(2016)}]{denig_review_2016}%
  \BibitemOpen
  \bibfield  {author} {\bibinfo {author} {\bibfnamefont {A.}~\bibnamefont
  {Denig}},\ }\bibfield  {title} {{\selectlanguage {english}\bibinfo {title}
  {Review of dark photon searches}},\ }\href
  {https://doi.org/10.1051/epjconf/201613001005} {\bibfield  {journal}
  {\bibinfo  {journal} {EPJ Web Conf.}\ }\textbf {\bibinfo {volume} {130}},\
  \bibinfo {pages} {01005} (\bibinfo {year} {2016})}\BibitemShut {NoStop}%
\bibitem [{\citenamefont {Curciarello}(2016)}]{curciarello_review_2016}%
  \BibitemOpen
  \bibfield  {author} {\bibinfo {author} {\bibfnamefont {F.}~\bibnamefont
  {Curciarello}},\ }\bibfield  {title} {{\selectlanguage {english}\bibinfo
  {title} {Review on {Dark} {Photon}}},\ }\href
  {https://doi.org/10.1051/epjconf/201611801008} {\bibfield  {journal}
  {\bibinfo  {journal} {EPJ Web of Conferences}\ }\textbf {\bibinfo {volume}
  {118}},\ \bibinfo {pages} {01008} (\bibinfo {year} {2016})}\BibitemShut
  {NoStop}%
\bibitem [{\citenamefont {Wojcik}\ and\ \citenamefont
  {Rizzo}(2022)}]{wojcik_forbidden_2022}%
  \BibitemOpen
  \bibfield  {author} {\bibinfo {author} {\bibfnamefont {G.~N.}\ \bibnamefont
  {Wojcik}}\ and\ \bibinfo {author} {\bibfnamefont {T.~G.}\ \bibnamefont
  {Rizzo}},\ }\bibfield  {title} {{\selectlanguage {english}\bibinfo {title}
  {Forbidden {Scalar} {Dark} {Matter} and {Dark} {Higgses}}},\ }\href
  {https://doi.org/10.1007/JHEP04(2022)033} {\bibfield  {journal} {\bibinfo
  {journal} {J. High Energ. Phys.}\ }\textbf {\bibinfo {volume}
  {2022}}\bibfield  {number} {\bibinfo  {number} { (4)},\ \bibinfo {pages}
  {33}},\ }\bibinfo {note} {arXiv:2109.07369 [hep-ph]}\BibitemShut {NoStop}%
\bibitem [{\citenamefont {Su}\ \emph {et~al.}(2022)\citenamefont {Su},
  \citenamefont {Wu},\ and\ \citenamefont {Zhu}}]{su_probing_2022}%
  \BibitemOpen
  \bibfield  {author} {\bibinfo {author} {\bibfnamefont {L.}~\bibnamefont
  {Su}}, \bibinfo {author} {\bibfnamefont {L.}~\bibnamefont {Wu}},\ and\
  \bibinfo {author} {\bibfnamefont {B.}~\bibnamefont {Zhu}},\ }\bibfield
  {title} {{\selectlanguage {english}\bibinfo {title} {Probing ultra-light dark
  photon from inverse {Compton}-like scattering}},\ }\href
  {https://doi.org/10.1103/PhysRevD.105.055021} {\bibfield  {journal} {\bibinfo
   {journal} {Phys. Rev. D}\ }\textbf {\bibinfo {volume} {105}},\ \bibinfo
  {pages} {055021} (\bibinfo {year} {2022})},\ \bibinfo {note}
  {arXiv:2105.06326 [astro-ph, physics:hep-ph]}\BibitemShut {NoStop}%
\bibitem [{\citenamefont {Redi}\ and\ \citenamefont
  {Tesi}(2022)}]{redi_dark_2022}%
  \BibitemOpen
  \bibfield  {author} {\bibinfo {author} {\bibfnamefont {M.}~\bibnamefont
  {Redi}}\ and\ \bibinfo {author} {\bibfnamefont {A.}~\bibnamefont {Tesi}},\
  }\href {http://arxiv.org/abs/2204.14274} {{\selectlanguage {english}\bibinfo
  {title} {Dark {Photon} {Dark} {Matter} without {Stueckelberg} {Mass}}}}
  (\bibinfo {year} {2022}),\ \bibinfo {note} {arXiv:2204.14274 [astro-ph,
  physics:hep-ph]}\BibitemShut {NoStop}%
\bibitem [{\citenamefont {Mondino}\ \emph {et~al.}(2021)\citenamefont
  {Mondino}, \citenamefont {Pospelov}, \citenamefont {Ruderman},\ and\
  \citenamefont {Slone}}]{mondino_dark_2021}%
  \BibitemOpen
  \bibfield  {author} {\bibinfo {author} {\bibfnamefont {C.}~\bibnamefont
  {Mondino}}, \bibinfo {author} {\bibfnamefont {M.}~\bibnamefont {Pospelov}},
  \bibinfo {author} {\bibfnamefont {J.~T.}\ \bibnamefont {Ruderman}},\ and\
  \bibinfo {author} {\bibfnamefont {O.}~\bibnamefont {Slone}},\ }\bibfield
  {title} {{\selectlanguage {english}\bibinfo {title} {Dark {Higgs} {Dark}
  {Matter}}},\ }\href {https://doi.org/10.1103/PhysRevD.103.035027} {\bibfield
  {journal} {\bibinfo  {journal} {Phys. Rev. D}\ }\textbf {\bibinfo {volume}
  {103}},\ \bibinfo {pages} {035027} (\bibinfo {year} {2021})},\ \bibinfo
  {note} {arXiv:2005.02397 [astro-ph, physics:hep-ph]}\BibitemShut {NoStop}%
\bibitem [{\citenamefont {Gorghetto}\ \emph {et~al.}(2022)\citenamefont
  {Gorghetto}, \citenamefont {Hardy}, \citenamefont {March-Russell},
  \citenamefont {Song},\ and\ \citenamefont {West}}]{gorghetto_dark_2022}%
  \BibitemOpen
  \bibfield  {author} {\bibinfo {author} {\bibfnamefont {M.}~\bibnamefont
  {Gorghetto}}, \bibinfo {author} {\bibfnamefont {E.}~\bibnamefont {Hardy}},
  \bibinfo {author} {\bibfnamefont {J.}~\bibnamefont {March-Russell}}, \bibinfo
  {author} {\bibfnamefont {N.}~\bibnamefont {Song}},\ and\ \bibinfo {author}
  {\bibfnamefont {S.~M.}\ \bibnamefont {West}},\ }\bibfield  {title}
  {{\selectlanguage {english}\bibinfo {title} {Dark {Photon} {Stars}:
  {Formation} and {Role} as {Dark} {Matter} {Substructure}}},\ }\href
  {https://doi.org/10.1088/1475-7516/2022/08/018} {\bibfield  {journal}
  {\bibinfo  {journal} {J. Cosmol. Astropart. Phys.}\ }\textbf {\bibinfo
  {volume} {2022}}\bibfield  {number} {\bibinfo  {number} { (08)},\ \bibinfo
  {pages} {018}},\ }\bibinfo {note} {arXiv:2203.10100 [astro-ph,
  physics:hep-ph]}\BibitemShut {NoStop}%
\bibitem [{\citenamefont {Marocco}(2021)}]{marocco_dark_2021}%
  \BibitemOpen
  \bibfield  {author} {\bibinfo {author} {\bibfnamefont {G.}~\bibnamefont
  {Marocco}},\ }\href {http://arxiv.org/abs/2110.02875} {{\selectlanguage
  {english}\bibinfo {title} {Dark photon limits from magnetic fields and
  astrophysical plasmas}}} (\bibinfo {year} {2021}),\ \bibinfo {note}
  {arXiv:2110.02875 [hep-ph]}\BibitemShut {NoStop}%
\bibitem [{\citenamefont {Fabbrichesi}\ \emph
  {et~al.}(2021{\natexlab{a}})\citenamefont {Fabbrichesi}, \citenamefont
  {Gabrielli},\ and\ \citenamefont {Lanfranchi}}]{fabbrichesi_dark_2021}%
  \BibitemOpen
  \bibfield  {author} {\bibinfo {author} {\bibfnamefont {M.}~\bibnamefont
  {Fabbrichesi}}, \bibinfo {author} {\bibfnamefont {E.}~\bibnamefont
  {Gabrielli}},\ and\ \bibinfo {author} {\bibfnamefont {G.}~\bibnamefont
  {Lanfranchi}},\ }\href {https://doi.org/10.1007/978-3-030-62519-1}
  {{\selectlanguage {english}\emph {\bibinfo {title} {The {Dark} {Photon}}}}}\
  (\bibinfo {year} {2021})\ \bibinfo {note} {arXiv:2005.01515 [hep-ex,
  physics:hep-ph]}\BibitemShut {NoStop}%
\bibitem [{\citenamefont {Fabbrichesi}\ \emph
  {et~al.}(2021{\natexlab{b}})\citenamefont {Fabbrichesi}, \citenamefont
  {Gabrielli},\ and\ \citenamefont {Lanfranchi}}]{fabbrichesi_physics_2021}%
  \BibitemOpen
  \bibfield  {author} {\bibinfo {author} {\bibfnamefont {M.}~\bibnamefont
  {Fabbrichesi}}, \bibinfo {author} {\bibfnamefont {E.}~\bibnamefont
  {Gabrielli}},\ and\ \bibinfo {author} {\bibfnamefont {G.}~\bibnamefont
  {Lanfranchi}},\ }\href {https://doi.org/10.1007/978-3-030-62519-1}
  {{\selectlanguage {english}\emph {\bibinfo {title} {The {Physics} of the
  {Dark} {Photon}: {A} {Primer}}}}},\ {SpringerBriefs} in {Physics}\ (\bibinfo
  {publisher} {Springer International Publishing},\ \bibinfo {address} {Cham},\
  \bibinfo {year} {2021})\BibitemShut {NoStop}%
\bibitem [{\citenamefont {Carroll}(2019)}]{carroll_spacetime_2019}%
  \BibitemOpen
  \bibfield  {author} {\bibinfo {author} {\bibfnamefont {S.~M.}\ \bibnamefont
  {Carroll}},\ }\href {https://doi.org/10.1017/9781108770385} {{\selectlanguage
  {english}\emph {\bibinfo {title} {Spacetime and {Geometry}: {An}
  {Introduction} to {General} {Relativity}}}}},\ \bibinfo {edition} {1st}\ ed.\
  (\bibinfo  {publisher} {Cambridge University Press},\ \bibinfo {year}
  {2019})\BibitemShut {NoStop}%
\bibitem [{\citenamefont {Jenkins}\ \emph {et~al.}(2013)\citenamefont
  {Jenkins}, \citenamefont {Manohar},\ and\ \citenamefont
  {Trott}}]{jenkins_gauge_2013}%
  \BibitemOpen
  \bibfield  {author} {\bibinfo {author} {\bibfnamefont {E.~E.}\ \bibnamefont
  {Jenkins}}, \bibinfo {author} {\bibfnamefont {A.~V.}\ \bibnamefont
  {Manohar}},\ and\ \bibinfo {author} {\bibfnamefont {M.}~\bibnamefont
  {Trott}},\ }\bibfield  {title} {{\selectlanguage {english}\bibinfo {title}
  {On {Gauge} {Invariance} and {Minimal} {Coupling}}},\ }\href
  {https://doi.org/10.1007/JHEP09(2013)063} {\bibfield  {journal} {\bibinfo
  {journal} {J. High Energ. Phys.}\ }\textbf {\bibinfo {volume}
  {2013}}\bibfield  {number} {\bibinfo  {number} { (9)},\ \bibinfo {pages}
  {63}},\ }\bibinfo {note} {arXiv:1305.0017 [hep-ph,
  physics:hep-th]}\BibitemShut {NoStop}%
\bibitem [{\citenamefont {Rubin}\ and\ \citenamefont
  {Ford}(1970)}]{rubin_rotation_1970}%
  \BibitemOpen
  \bibfield  {author} {\bibinfo {author} {\bibfnamefont {V.~C.}\ \bibnamefont
  {Rubin}}\ and\ \bibinfo {author} {\bibfnamefont {W.~K.}\ \bibnamefont {Ford},
  \bibfnamefont {Jr.}},\ }\bibfield  {title} {{\selectlanguage
  {english}\bibinfo {title} {Rotation of the {Andromeda} {Nebula} from a
  {Spectroscopic} {Survey} of {Emission} {Regions}}},\ }\href
  {https://doi.org/10.1086/150317} {\bibfield  {journal} {\bibinfo  {journal}
  {ApJ}\ }\textbf {\bibinfo {volume} {159}},\ \bibinfo {pages} {379} (\bibinfo
  {year} {1970})}\BibitemShut {NoStop}%
\bibitem [{\citenamefont {Anderson}(1981)}]{anderson_principle_1981}%
  \BibitemOpen
  \bibfield  {author} {\bibinfo {author} {\bibfnamefont {I.~M.}\ \bibnamefont
  {Anderson}},\ }\bibfield  {title} {{\selectlanguage {english}\bibinfo {title}
  {The principle of minimal gravitational coupling}},\ }\href
  {https://doi.org/10.1007/BF00256383} {\bibfield  {journal} {\bibinfo
  {journal} {Arch. Rational Mech. Anal.}\ }\textbf {\bibinfo {volume} {75}},\
  \bibinfo {pages} {349} (\bibinfo {year} {1981})}\BibitemShut {NoStop}%
\bibitem [{\citenamefont {Melo}(2017)}]{melo_higgs_2017}%
  \BibitemOpen
  \bibfield  {author} {\bibinfo {author} {\bibfnamefont {I.}~\bibnamefont
  {Melo}},\ }\bibfield  {title} {{\selectlanguage {english}\bibinfo {title}
  {Higgs potential and fundamental physics}},\ }\href
  {https://doi.org/10.1088/1361-6404/aa8c3d} {\bibfield  {journal} {\bibinfo
  {journal} {Eur. J. Phys.}\ }\textbf {\bibinfo {volume} {38}},\ \bibinfo
  {pages} {065404} (\bibinfo {year} {2017})},\ \bibinfo {note}
  {arXiv:1911.08893 [hep-ph, physics:physics]}\BibitemShut {NoStop}%
\bibitem [{\citenamefont {Bertone}\ and\ \citenamefont
  {Tait}(2018)}]{bertone_new_2018}%
  \BibitemOpen
  \bibfield  {author} {\bibinfo {author} {\bibfnamefont {G.}~\bibnamefont
  {Bertone}}\ and\ \bibinfo {author} {\bibfnamefont {T.~M.~P.}\ \bibnamefont
  {Tait}},\ }\bibfield  {title} {{\selectlanguage {english}\bibinfo {title} {A
  {New} {Era} in the {Quest} for {Dark} {Matter}}},\ }\href
  {https://doi.org/10.1038/s41586-018-0542-z} {\bibfield  {journal} {\bibinfo
  {journal} {Nature}\ }\textbf {\bibinfo {volume} {562}},\ \bibinfo {pages}
  {51} (\bibinfo {year} {2018})},\ \bibinfo {note} {arXiv:1810.01668 [astro-ph,
  physics:hep-ph]}\BibitemShut {NoStop}%
\bibitem [{\citenamefont {Bergstr{\"o}m}(2012)}]{bergstrom_dark_2012}%
  \BibitemOpen
  \bibfield  {author} {\bibinfo {author} {\bibfnamefont {L.}~\bibnamefont
  {Bergstr{\"o}m}},\ }\bibfield  {title} {{\selectlanguage {english}\bibinfo
  {title} {Dark {Matter} {Evidence}, {Particle} {Physics} {Candidates} and
  {Detection} {Methods}}},\ }\href {https://doi.org/10.1002/andp.201200116}
  {\bibfield  {journal} {\bibinfo  {journal} {Ann. Phys.}\ }\textbf {\bibinfo
  {volume} {524}},\ \bibinfo {pages} {479} (\bibinfo {year} {2012})},\ \bibinfo
  {note} {arXiv:1205.4882 [astro-ph, physics:hep-ph]}\BibitemShut {NoStop}%
\bibitem [{\citenamefont {Lewis}(2009)}]{lewis_explicit_2009}%
  \BibitemOpen
  \bibfield  {author} {\bibinfo {author} {\bibfnamefont {C.~L.}\ \bibnamefont
  {Lewis}},\ }\bibfield  {title} {{\selectlanguage {english}\bibinfo {title}
  {Explicit gauge covariant {Euler}{\textendash}{Lagrange} equation}},\ }\href
  {https://doi.org/10.1119/1.3153503} {\bibfield  {journal} {\bibinfo
  {journal} {American Journal of Physics}\ }\textbf {\bibinfo {volume} {77}},\
  \bibinfo {pages} {839} (\bibinfo {year} {2009})}\BibitemShut {NoStop}%
\bibitem [{\citenamefont {Keshet}\ and\ \citenamefont
  {Gurwich}(2018)}]{keshet_fermi_2018}%
  \BibitemOpen
  \bibfield  {author} {\bibinfo {author} {\bibfnamefont {U.}~\bibnamefont
  {Keshet}}\ and\ \bibinfo {author} {\bibfnamefont {I.}~\bibnamefont
  {Gurwich}},\ }\bibfield  {title} {{\selectlanguage {english}\bibinfo {title}
  {Fermi bubbles: high latitude {X}-ray supersonic shell}},\ }\href
  {https://doi.org/10.1093/mnras/sty1533} {\bibfield  {journal} {\bibinfo
  {journal} {Monthly Notices of the Royal Astronomical Society}\ }\textbf
  {\bibinfo {volume} {480}},\ \bibinfo {pages} {223} (\bibinfo {year}
  {2018})},\ \bibinfo {note} {arXiv:1704.05070 [astro-ph]}\BibitemShut
  {NoStop}%
\bibitem [{\citenamefont {Sanchis-Gual}\ \emph {et~al.}(2017)\citenamefont
  {Sanchis-Gual}, \citenamefont {Herdeiro}, \citenamefont {Radu}, \citenamefont
  {Degollado},\ and\ \citenamefont {Font}}]{sanchis-gual_numerical_2017}%
  \BibitemOpen
  \bibfield  {author} {\bibinfo {author} {\bibfnamefont {N.}~\bibnamefont
  {Sanchis-Gual}}, \bibinfo {author} {\bibfnamefont {C.}~\bibnamefont
  {Herdeiro}}, \bibinfo {author} {\bibfnamefont {E.}~\bibnamefont {Radu}},
  \bibinfo {author} {\bibfnamefont {J.~C.}\ \bibnamefont {Degollado}},\ and\
  \bibinfo {author} {\bibfnamefont {J.~A.}\ \bibnamefont {Font}},\ }\bibfield
  {title} {{\selectlanguage {english}\bibinfo {title} {Numerical evolutions of
  spherical {Proca} stars}},\ }\href
  {https://doi.org/10.1103/PhysRevD.95.104028} {\bibfield  {journal} {\bibinfo
  {journal} {Phys. Rev. D}\ }\textbf {\bibinfo {volume} {95}},\ \bibinfo
  {pages} {104028} (\bibinfo {year} {2017})},\ \bibinfo {note}
  {arXiv:1702.04532 [gr-qc]}\BibitemShut {NoStop}%
\end{thebibliography}%
